\documentclass[twocolumn,aps,showpacs,prb,tightenlines,amsmath,amssymb]{revtex4}
\usepackage{graphicx}
\usepackage{amssymb}
\usepackage{slashed}
\usepackage{dcolumn}
\usepackage{amsmath}
\usepackage{bm}
\usepackage{colordvi}
\usepackage{mathrsfs}
\makeatletter

\newcommand{\Rmnum}[1]{\expandafter\@slowromancap\romannumeral #1@}
\makeatother

\begin{document}

\title{Impurity scattering in superconductors revisited}   

\author{F. Yang}
\email{yfgq@mail.ustc.edu.cn.}

\affiliation{Hefei National Research Center for Physical Sciences at the Microscale, Department of Physics, and CAS Key Laboratory of Strongly-Coupled
Quantum Matter Physics, University of Science and Technology of China, Hefei,
Anhui, 230026, China}

\author{M. W. Wu}
\email{mwwu@ustc.edu.cn.}

\affiliation{Hefei National Research Center for Physical Sciences at the Microscale, Department of Physics, and CAS Key Laboratory of Strongly-Coupled
Quantum Matter Physics, University of Science and Technology of China, Hefei,
Anhui, 230026, China}

\date{\today}

\begin{abstract}

  The diagrammatic formalism and transport equation are conventionally considered as separate but complementary techniques to tackle the impurity scattering effect.  To compare with the previous studies from the gauge-invariant kinetic equation approach [F. Yang and M. W. Wu, Phys. Rev. B {\bf 98}, 094507 (2018); {\bf 102}, 144508 (2020)],  we analytically perform a diagrammatic formulation of the impurity scattering in superconductors, with both transport and collective Higgs mode studied, in order to fill the long missing calculation of the Kubo current-current correlation in superconductors with impurity scattering and  resolve the controversy (whether the impurity scattering can lead to the damping of Higgs
  mode) between gauge-invariant kinetic equation and Eilenberger equation. For transport behavior, through a special unitary transformation that is equivalent to the Wilson-line technique for the diamagnetic response, we derive the Meissner-supercurrent vertex. Then, by formulating the supercurrent-supercurrent correlation with Born and vertex corrections from impurity scattering, we recover the previously revealed microscopic momentum-relaxation rate of superfluid by gauge-invariant kinetic equation. This rate is finite only when the superconducting velocity is larger than a threshold, at which the normal fluid emerges and causes the friction with the superfluid current, similar to the Landau's superfluid theory of liquid helium. This derivation also provides a physical understanding of the relaxation-time approximation in the previous diagrammatic formulation in the literature, which leads to the friction resistance of the Meissner supercurrent. For the collective Higgs mode, we calculate the amplitude-amplitude correlation with Born and vertex corrections from impurity scattering. The vertex correction, which only emerges at nonequilibrium case, leads to a Higgs-mode damping, whereas the Born correction that is equivalent to equilibrium self-energy makes no contribution due to the Anderson theorem. This induced damping agrees with the analysis through Heisenberg equation of motion and is also exactly same as the one obtained from gauge-invariant kinetic equation.  

\end{abstract}


\maketitle 

\section{Introduction}

The impurity scattering effect has attracted much attention in the field of superconductivity. On one hand, the stationary magnetic-flux expulsion due to the generated diamagnetic supercurrent (Meissner effect)\cite{Meissner,London}, as well as the low-frequency optical conductivity described by phenomenological two-fluid model\cite{TFM1,TFM2}, are characteristic transport properties of superconductors, among which elucidating the impurity scattering effect is essential to understand the superconductivity/resistivity phenomena. On the other hand, recently, inspired by nonlinear optical experiments in THz regime\cite{NL1,NL2,NL3,NL4,DHM1,DHM2,DHM3}, a great deal of efforts have been devoted to the collective gapful Higgs mode, which describes the amplitude fluctuation of the superconducting order parameter\cite{Am0,OD1,OD2,OD3,Am12,Am3,Am4,Am5,Am6,symmetry}. Being charge neutral, this collective excitation does not manifest itself in the linear optical response, but can be generated in the second-order one at clean limit\cite{EPM}, leading to an experimentally observable fluctuation of superfluid density\cite{EPM}. The damping mechanism of the Higgs mode after excitation has then stimulated a lot of interest\cite{DOHM,DHM2}.

Theoretically, two kinds of schemes have been developed in the literature to formulate the impurity scattering effect, including the diagrammatic formalism and transport equation, which are conventionally considered as separate but complementary techniques as demonstrated in normal metals\cite{MP}. Nevertheless, in superconductors, the relationship of two techniques has not been well developed in the literature for decades. 

The formulation within the diagrammatic formalism requires the inevitable calculation of the vertex correction by impurity scattering\cite{G0,G1}, which becomes hard to tackle in superconductors. Specifically, it is established\cite{NSL10} that the superconductors with the small (large) mean free path $l$ in comparison with the skin depth $\delta$ lie in the normal (anomalous) skin-effect region and exhibit the London-type/local (Pippard-type/non-local) electromagnetic response. The linear electromagnetic responses of superconductors in the anomalous- and normal-skin-effect regions were first discussed by Mattis and Bardeen\cite{MB} as well as Abrikosov and Gorkov\cite{G1},  based on the current-current correlation with the impurity scattering. To handle the scattering effect, the Mattis-Bardeen theory introduces a phenomenological constant scattering factor\cite{MB}, which is similar to the relaxation-time approximation. Whereas Abrikosov and Gorkov\cite{G1} applied an approximation that assumes an isotropic Green function in consideration of a dirty case ($l\ll\xi$, with $\xi$ being the coherence length), in order to integrate over the momentum variable of pairing electrons to simplify the vertex-correction calculation. Both approximations then drop out the microscopic scattering process. Interestingly, both Mattis-Bardeen\cite{MB,MBo} and Abrikosov-Gorkov\cite{G1} theoretical descriptions in the diamagnetic response derive a penetration depth $\lambda=\lambda_c\sqrt{\xi/l}$ at dirty limit, with $\lambda_c$ being the clean-limit result. By using the relaxation-time approximation, this dependence was later phenomenologically extended by Tinkham\cite{Ba1} to a general form $\lambda=\lambda_c\sqrt{1+\xi/l}$ between clean and dirty cases, in good agreement with the experiments\cite{E1,E2,E3,E4,E5}. Nevertheless, as a direct consequence of this dependence that is derived from the current-current correlation, the Meissner supercurrent, which should be non-viscous, experiences a friction resistance by scattering. The physical origin of this resistance becomes untraceable due to the absent microscopic scattering process. Moreover, these theoretical descriptions also fail to recover the two-fluid model, which requires a microscopic distinction of the pairing (superfluid) and unpairing (normal fluid) electrons\cite{TFM1,FF1,FF2}. In contrast to the transport behavior, the diagrammatic formulation of the impurity scattering effect on Higgs mode remains stagnant so far. While the calculation of the amplitude-amplitude correlation at clean case successfully gives the Higgs-mode energy spectrum in long-wave limit\cite{Cea1,Cea2,aa1,Cea3,aa2}, it is complicated to formulate the corresponding vertex correction by the impurity scattering.

The transport-equation approach with microscopic scattering can naturally contain and easily handle the calculation of the vertex correction by scattering, as demonstrated in normal metals\cite{Q1,MP}. In superconductors, three kinds of transport equations that construct the microscopic scattering have been developed in the literature, including semi-classical Boltzmann equation of quasiparticles\cite{Ba3,Bol,Ba5}, quasiclassical Eilenberger equation\cite{Eilen,Ba7,Ba8,Ba20,Eilen1}  as well as gauge-invariant kinetic equation (GIKE)\cite{GIKE1,GIKE2,GIKE3,GIKE4}. The semi-classical Boltzmann equation as an early stage of works only includes the quasiparticle dynamics but fails to contain the superfluid dynamics\cite{Ba3,Bol,Ba5}.

The Eilenberger equation\cite{Eilen,Ba20,Eilen1} is derived from the basic Gorkov equation\cite{G1} of $\tau_3$-Green function {\small $G_3(x,x')=-i\tau_3\langle{\hat T}\psi(x)\psi^{\dagger}(x')\rangle$}, through the quasiclassical approximation which performs an integration over kinetic-energy variable. Here, $\tau_i$ denotes the Pauli matrices in Nambu space. This approach at free case can successfully describe the Higgs-mode energy spectrum\cite{Silaev} and discuss the topics like proximity effect in multilayer junctions\cite{Ba7,Ba8} as well as vortex dynamics\cite{EV1,EV2,EV3,EV4} and unconventional superconductivity\cite{EU1,EU2,EU3,EU4}. While concerning the electromagnetic response, the gauge invariance is lost during the derivation\cite{EG}, leading to incomplete electromagnetic effect. As a consequence, the Eilenberger equation only keeps the drive effect of vector potential\cite{EF}, making it well tailored to handle the diamagnetic response (i.e., derive the Ginzburg-Landau equation as well as Meissner supercurrent\cite{Ba20}) and gives a finite Higgs-mode generation in the second optical response at clean limit\cite{EPM,Silaev1}. But the drive effect by scalar potential and all density-related electromagnetic effects are generically dropped out\cite{EPM,EF}.

Focusing on the scattering effect, the Eilenberger equation contains the specific quasiclassical microscopic scattering integral\cite{Eilen,Ba7,Ba8,Ba20,Eilen1,Silaev}. In the diamagnetic response, the derived supercurrent from this approach also experiences a friction resistance\cite{Silaev,Eilen1,Eilen}. Particularly, in the Usadel equation\cite{Usadel}, which is a dirty-limit case of Eilenberger equation, the induced supercurrent is directly proportional to the diffusive coefficient, in consistency with the Mattis-Bardeen\cite{MB,MBo} and Abrikosov-Gorkov\cite{G1} theoretical descriptions mentioned above. Nevertheless, elucidating the origin of this friction resistance has long been overlooked. As for the collective excitation, with impurities, it is reported\cite{Silaev,Silaev0} that the derived Higgs-mode energy spectrum in Eilenberger equation is free from the scattering influence, i.e., the impurity scattering does not cause the damping of Higgs mode. In one view in the literature, the Higgs mode as the gap fluctuation is insensitive to disorder, as the Anderson theorem\cite{ISE} reveals a vanishing renormalization by impurity self-energy on equilibrium $s$-wave gap\cite{ISE1,ISE2,ISE3,ISE4} in consideration of the time-reversal-partner pairing. Very recently, this
 viewpoint is challenged\cite{GIKE3}. The key point lies at the fact that the Higgs mode is a nonequilibrium excitation which breaks the time-translational symmetry. Thus, applying the Anderson theorem to nonequilibrium case is unsuitable. In this circumstance, considering the fact that the Higgs-mode excitation $\delta|\Delta|\tau_1$ and electron-impurity interaction $V({\bf r})\tau_3$ are non-commutative in Nambu space, one immediately concludes that the nonequilibrium Higgs mode experiences a finite impurity influence according to Heisenberg equation of motion. This analysis is then in sharp contrast to the derivation from the Eilenberger equation\cite{Silaev} mentioned above.

The GIKE\cite{GIKE1,GIKE2} is derived from the Gorkov equation\cite{G1} of $\tau_0$-Green function {\small $G_0(x,x')=-i\langle{\hat T}\psi(x)\psi^{\dagger}(x')\rangle$} within equal-time scheme\cite{GQ2,GQ3}. To retain the gauge invariance, the gauge-invariant $\tau_0$-Green function is constructed through the Wilson line\cite{Wilson}. Then, the complete electromagnetic effects are included\cite{GIKE1} and the charge conservation is naturally satisfied\cite{GIKE2}, making this approach capable of formulating both magnetic and optical responses in linear and nonlinear regimes. The well-known clean-limit results, such as Ginzburg-Landau equation and Meissner supercurrent in the diamagnetic response and the low-frequency optical conductivity captured by the two-fluid model\cite{GIKE1} as well as the linear electromagnetic responses of the collective phase and Higgs modes\cite{GIKE2}, can be directly derived from this microscopic approach. Very recently, in the second optical response at clean limit, the derived finite Higgs-mode generation and vanishing charge-density fluctuation from GIKE\cite{GIKE2} are exactly recovered from the basic path-integral approach\cite{EPM}.

Thanks to the equal-time scheme\cite{GQ2,GQ3}, the microscopic scattering in superconductors, which is hard to tackle within the diagrammatic formalism, becomes easy to handle within the GIKE. From this approach, not only
the previously revealed phenomenological dependence of the penetration depth on mean-free path by Tinkham is recovered\cite{GIKE1}, but also the disorder-induced damping of Higgs mode is revealed for the first time\cite{GIKE3}. Specifically, it is analytically demonstrated\cite{GIKE1} that the generated Meissner supercurrent in diamagnetic response becomes viscous only when the superconducting velocity is larger than a threshold at which the normal fluid emerges, similar to the Landau's theory for the emerged fluid viscosity in bosonic liquid helium at larger velocity\cite{Landau}. The emergence of the viscous superfluid in superconductors arises from the friction between normal-fluid and superfluid currents due to the microscopic scattering\cite{GIKE1}. A three-fluid model consisting of normal fluid as well as viscous and non-viscous superfluids is then proposed\cite{GIKE1,GIKE4}.  As for the damping of Higgs mode, it is found\cite{GIKE3} that the impurity scattering leads to a fast exponential decay, which arises from the non-commutation relation between Higgs mode and electron-impurity interaction.
This damping then agrees with the analysis through the Heisenberg equation of motion mentioned above, but is in contrast to the previous derivation from Eilenberger equation\cite{Silaev}.

In the present work, to achieve a separate but complementary technique to compare with GIKE, we try to apply the diagrammatic formulation of the impurity scattering in superconductors, in order to fill the long missing calculation of the Kubo current-current correlation in superconductors with the impurity scattering in the textbook and resolve the controversy (whether the impurity scattering can lead to the damping of Higgs mode) between GIKE\cite{GIKE3} and Eilenberger equation\cite{Silaev} mentioned above. Specifically, for transport behavior in the diamagnetic response, because of the Meissner effect\cite{Meissner,London}, it is shown that the density vertex $\tau_3$\cite{Ba0} in the conventional kinematical momentum operator ${\hat {\bf p}}-e{\bf A}\tau_3$ leads to a non-gauge-invariant current after the scattering treatment/correction on the current-current correlation. To eliminate this unphysical current, we apply a special unitary transformation that is equivalent to the Wilson-line technique for the diamagnetic response, and obtain the Meissner-supercurrent vertex. Then, by further calculating the supercurrent-supercurrent correlation with Born and vertex corrections from the impurity scattering,  the microscopic momentum-relaxation rate of superfluid, which is exactly same as the one from GIKE\cite{GIKE1,GIKE5}, is derived. This rate becomes finite only when the superconducting velocity is larger than a threshold, at which the normal fluid emerges and causes the friction with superfluid current. Then, the three-fluid model proposed in Ref.~\onlinecite{GIKE1} is recovered. Moreover, this derivation also provides a physical understanding of the relaxation-time approximation in the previous diagrammatic formulation\cite{MB,MBo,Ba1}, which leads to the friction resistance of the Meissner supercurrent as mentioned above. Furthermore, through the Wilson-line technique, a gauge-invariant Hamiltonian that explicitly distinguishes the Meissner effect and electric-field drive effect as well as the Josephson  voltage effect, is proposed. 

As for the collective Higgs mode, we perform an analytical calculation of the amplitude-amplitude correlation with the Born and vertex corrections from impurity scattering. It is found that the vertex correction leads to a fast exponential damping of Higgs mode, whereas the Born correction that is equivalent to equilibrium impurity self-energy makes no contribution because of the Anderson theorem\cite{ISE,ISE1,ISE2,ISE3,ISE4}. This induced damping by impurity scattering is exactly same as the one obtained from GIKE and agrees with analysis through Heisenberg equation of motion mentioned above, in contrast to the previous derivation by Eilenberger equation\cite{Silaev}. The revealed lifetime of Higgs mode by impurity scattering provides a possible origin for the experimentally observed broadening of the resonance signal\cite{NL3,NL4} as well as the damping after optical excitation\cite{NL1,NL2} of the Higgs mode. In addition, as pointed out in Ref.~\onlinecite{GIKE3}, the damping by impurity can cause a phase shift in the optical signal of Higgs mode, which exhibits a $\pi$-jump at the resonance frequency and hence provides a very clear feature for the further experimental detection.

\section{Model}

In this section, we first introduce the Hamiltonian and action of superconductors in the presence of the superconducting momentum. Then, based on the basic path-integral approach, we present the diagrammatic formalism to investigate the scattering effects on nonequilibrium property in superconductors.  

\subsection{Hamiltonian and action}
\label{secHA}

It is well known in superconductors that in the stationary magnetic response with a vector potential ${\bf A}$, a supercurrent is driven by superconducting momentum ${\bf p}_s=-e{\bf A}$\cite{G1}. The Bogoliubov-de Gennes Hamiltonian of the conventional $s$-wave superconducting states in the presence of the superconducting momentum reads\cite{G1}
\begin{equation}
  H\!=\!\!{\int}{d{\bf x}}\psi^{\dagger}(x)[\xi_{{\hat{\bf p}}+{\bf p}_s\tau_3}\tau_3+\Delta_0\tau_1+V(x)\tau_3]\psi(x),\label{BdG}
\end{equation}
where $\psi(x)=[\psi_{\uparrow}(x),\psi^{\dagger}_{\downarrow}(x)]^T$ represents the field operator in Nambu space with $x=(x_0,{\bf x})$ being the space-time four-vector; the momentum operator ${\hat {\bf p}}=-i\hbar{\bm \nabla}$; $\xi_{\hat {\bf p}}={{\hat {\bf p}}^2}/(2m)-\mu$ with $m$ denoting the effective mass and $\mu$ being the chemical potential; $\Delta_0$ and $V(x)$ denote the equilibrium gap and impurity potential, respectively.  

Based on the Hamiltonian above, the action of superconductors after Hubbard-Stratonovich transformation is written as\cite{Ba0}
\begin{eqnarray}
\!\!\!\!\!S&=&\int{dx}\bigg\{\sum_{s=\uparrow,\downarrow}\!\!\psi^*_s(x)[i\partial_{x_0}\!-\!\xi_{\hat {\bf p}+{\bf p}_s}\!-\!V(x)]\psi_s(x)\nonumber\\
  &&\mbox{}-\!\psi^{\dagger}(x)\Delta_0\tau_1\psi(x)\!-\!\frac{\Delta_0^2}{g}\bigg\}. \label{SC}
\end{eqnarray}
which in Nambu space becomes
\begin{eqnarray}
S&=&\int{dx}\bigg\{\psi^{\dagger}(x)\big[G_0^{-1}({\hat {p}})\!-\!V(x)\tau_3\big]\psi(x)\!-\!\frac{\eta_fp_s^2}{2m}\nonumber\\
&&\mbox{}-\!\eta_fV(x)\!-\!\frac{|\Delta(x)|^2}{g}\bigg\}. \label{BdGaction}
\end{eqnarray}
Here, $g$ denotes the BCS pairing potential and $\eta_f=\sum_{\bf k}1$ emerges because of the anti-commutation of Fermi field; the Green function $G^{-1}_0({\hat p})=i\partial_{x_0}-{\bf p}_s\cdot{\bf v}_{\bf {\hat p}}-(\xi_{\hat {\bf p}}+\frac{p_s^2}{2m})\tau_3-{\Delta}_0\tau_1$ with ${\bf v}_{\bf {\hat p}}={\hat {\bf p}}/m$ standing for the group velocity.

The Fourier component of the Green function in Matsubara representation is given by\cite{G1}
\begin{equation}\label{Greenfunction}
G_0(p)=\frac{ip_n\!-\!{\bf p}_s\!\cdot\!{\bf v_k}\!+\!\xi_{\bf k}\tau_3\!+\!p_s^2\tau_3/(2m)\!+\!\Delta_0\tau_1}{(ip_n-E_{\bf k}^+)(ip_n-E_{\bf k}^-)},  
\end{equation}
where the four-vector momentum $p=(p_n,{\bf k})$ with $p_n=(2n+1)\pi{T}$ being the Matsubara frequency; the quasiparticle energy spectra read
\begin{equation}
  E_{\bf k}^{\pm}={\bf p}_s\cdot{\bf v_k}\pm{E_{\bf k}},
\end{equation}
with $E_{\bf k}=\sqrt{[\xi_{\bf k}+p_s^2/(2m)]^2+\Delta_0^2}$.

It is noted that the term ${\bf p}_s\cdot{{\bf v_k}}$ in the equilibrium Green function denotes the Doppler shift\cite{FF4,FF5,FF6,GIKE1,FF8,FF9,GIKE5}, which causes a tilted quasiparticle energy spectrum and hence markedly influences the superconducting anomalous correlation. Specifically, the gap equation reads\cite{G1} 
\begin{equation}\label{G11}
  {\rm {\bar Tr}}[G_0(p)\tau_1]=\sum_{\bf k}{2\Delta_0}F_{\bf k}=-{\Delta_0}/{g},  
\end{equation}
where the anomalous correlation $F_{\bf k}$ is written as
\begin{equation}
F_{\bf k}=\frac{f(E_{\bf k}^+)-f(E_{\bf k}^-)}{2E_{\bf k}}.  
\end{equation}
In momentum space, the anomalous correlation $F_{\bf k}$ vanishes in regions with $|{\bf p}_s\cdot{{\bf v_k}}|>E_{\bf k}$ where the quasielectron energy $E_{\bf k}^+<0$ or quasihole energy $E_{\bf k}^->0$, but remains finite in the regions with $|{\bf p}_s\cdot{{\bf v_k}}|<E_{\bf k}$.  Following the idea of Fulde-Ferrell-Larkin-Ovchinnikov state in conventional superconductors\cite{FF1,FF2}, regions with nonzero and vanishing anomalous correlation are referred to as the pairing and unpairing regions, respectively. Particles in pairing region contribute to the gap as superfluid, whereas particles in unpairing region no longer participate in the pairing and behave like normal ones, leading to the emergence of normal fluid\cite{FF1,GIKE1}. However, in the discussion of the scattering effect, the essential Doppler shift term was approximately neglected in the previous works\cite{MB,MBo,G1} and has long been overlooked in the literature. In the present work, we sublate this approximation by keeping the Doppler-shift term in the Green function.  

\subsection{Diagrammatic formalism}
\label{sec-df}

In this part, through the path-integral approach, we present the diagrammatic formalism to calculate the impurity scattering on nonequilibrium properties. For transport behavior in the diamagnetic response of superconductors, a supercurrent ${\bf j}$ is driven by a superconducting momentum ${\bf p}_s=-e{\bf A}$. At clean limit, ${\bf j}=-e^2n_s{\bf A}/m$ with $n_s$ being the superfluid density, and the penetration depth then reads {${\lambda_c}=\sqrt{m/(4{\pi}e^2n_s)}$}\cite{G1}. Nevertheless, with impurities, considering the friction resistance of supercurrent mentioned in the introduction as well as the role of the Doppler shift mentioned above, a self-consistent equation of motion of the superconducting current is required. In this circumstance, following the technique of applying the test charge in the Coulomb screening calculation\cite{MP}, we consider a test nonequilibrium variation $\delta{\bf p}_s(x)=-e\delta{\bf A}$ on top of the uniform ${\bf p}_s=-e{\bf A}$, which leads to a nonequilibrium variation of the superconducting current [${\bf j}\rightarrow{\bf j}+\delta{\bf j}(x)$].
    Then, by deriving the linear response of $\delta{\bf j}$ to $\delta{\bf p}_s$, one equivalently obtains the self-consistent equation of motion of the superconducting current. 
    
As for the Higgs mode [i.e., nonequilibrium gap fluctuation $\delta|\Delta|(x)$],  its equation of motion $(\partial_t^2-\omega_H^2)\delta|\Delta|=0$ at clean case, showing a gapful energy spectrum $\omega_H=2\Delta_0$ in long-wave limit, has been revealed by various theoretical approaches in the literature\cite{Silaev,aa2,GIKE2,GIKE4,GIKE3,Cea1,Cea2,aa1,Cea3,Am0,Am12,EPM,Am3,Am4,Am5,Am6,symmetry}. To discuss the damping, one also needs to derive the equation of motion of $\delta|\Delta|$ in the presence of the scattering.

Therefore, we consider a general self-energy $\Sigma_{\delta}(x,p)$ by nonequilibrium variation. The action including this nonequilibrium self-energy is written as 
\begin{eqnarray}
  S&\!\!\!\!=\!\!\!&\!\!\int\!\!{dx}\psi^{\dagger}(x)[G_0^{-1}({\hat p})\!-\!{\Sigma}_{\delta}(x,p)\!-\!V(x)\tau_3]\psi(x)\nonumber\\
&&\mbox{}\!\!-\!\!\int\!{dx}\Big[\eta_f\Sigma_{\delta3}\!+\!\frac{\eta_fp_s^2}{2m}\!+\!\eta_fV(x)\!+\!\frac{(\Delta_0\!+\!\delta|\Delta|)^2}{g}\Big],~~~
\end{eqnarray}
where $\Sigma_{\delta{i}}$ denotes the $\tau_i$ component of $\Sigma_{\delta}$. Through the integration over the Fermi field within the path-integral approach, one obtains the effective action: 
\begin{eqnarray}
  {S}&\!\!=&\!\!\!\int\!{dx}\Big[{\rm {\bar Tr}}\ln[G_0^{-1}\!-\!\Sigma_{\delta}\!-\!V\tau_3]\!-\!\frac{(\Delta_0+\delta|\Delta|)^2}{g}\!-\!\eta_f\nonumber\\
  &&\mbox{}\!\times\Big(\Sigma_{\delta3}\!+\!V\!+\!\frac{\eta_fp_s^2}{2m}\Big)\Big]\nonumber\\
  &\!\!=&\!\!\!\int\!{dx}\Big[{\rm {\bar Tr}}\ln{G_0^{-1}}\!-\!\eta_f\frac{p_s^2}{2m}\!-\!nV\!-\!\frac{(\Delta_0\!+\!\delta|\Delta|)^2}{g}\!-\!\eta_f\Sigma_{\delta3}\nonumber\\
  &&\mbox{}\!-\!{\rm{\bar Tr}}(G_0\Sigma_{\delta})\!-\!\sum_{n=2}^{\infty}\frac{1}{n}{\rm{\bar Tr}}\{[G_0(\Sigma_{\delta}\!+\!V\tau_3)]^n\}\Big],
\end{eqnarray}
where we have used ${\rm {\bar Tr}}[G_0\tau_3]+\eta_f=n$ (refer to Appendix~\ref{a-cd}) with $n$ denoting the charge density.

The equilibrium part in the effective action above reads 
\begin{eqnarray}
  S_0\!\!&=&\!\!\!\int\!{dx}\Big\{\sum_{p_n,{\bf k}}\ln{(ip_n\!-\!E_{\bf k}^+)(ip_n\!-\!E_{\bf k}^-)}\!-\!\Big(\frac{\eta_fp_s^2}{2m}\!+\!\frac{\Delta_0^2}{g}\Big)\nonumber\\
  &&\mbox{}\!\!\!-\sum_{n=2}^{\infty}\frac{1}{n}{\rm {\bar Tr}}[(G_0V\tau_3)^n]\Big\}. \label{S0}
\end{eqnarray}
It is noted that the last term on the right hand side of above equation denotes the equilibrium impurity self-energy\cite{ISE1,ISE2,ISE3,ISE4}, which in principle can cause renormalization on the equilibrium parameters, such as effective mass, chemical potential (charge density), superconducting momentum as well as gap\cite{MP}.

\begin{figure}[htb]
  {\includegraphics[width=8.7cm]{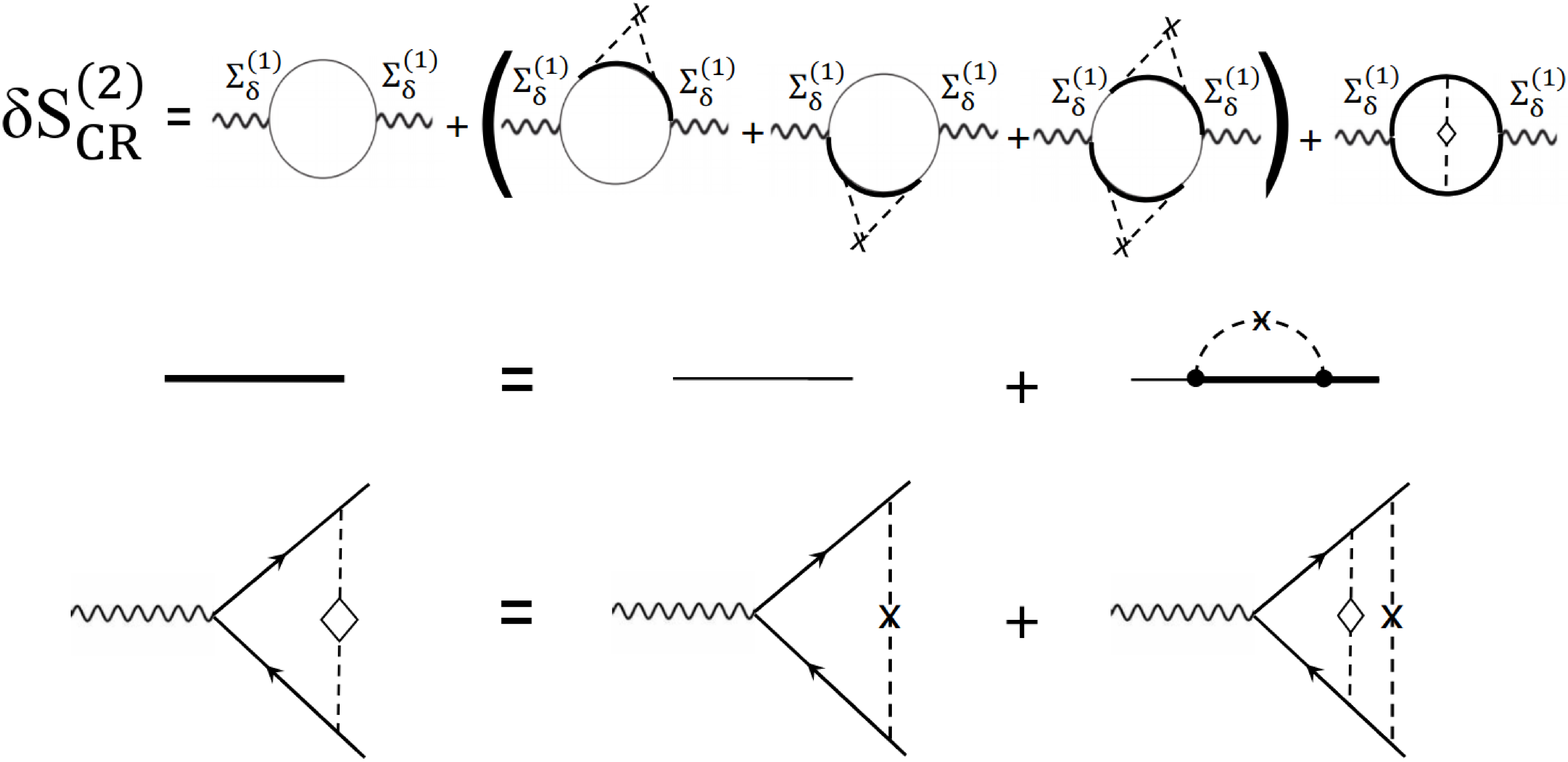}}
  \caption{Diagrammatic formalism for $\delta{S}^{(2)}_{\rm CR}$ (i.e., $\Sigma_{\delta}^{(1)}$-$\Sigma_{\delta}^{(1)}$ correlation). On the right-hand side of the equation of $\delta{S}^{(2)}_{\rm CR}$, the first diagram denotes the bare $\Sigma_{\delta}^{(1)}$-$\Sigma_{\delta}^{(1)}$ correlation; the second and third diagrams represent the Born and vertex corrections by impurity scattering, respectively. In the figure,  the dashed lines with cross and diamond represent the impurity interaction and ladder diagram of impurity scattering, respectively; the wavy line is associated with the nonequilibrium self-energy $\Sigma_{\delta}^{(1)}$; the thin and thick solid lines denote the bare and renormalized Green functions, respectively. }    
\label{figyw1}
\end{figure}

\begin{figure}[htb]
  {\includegraphics[width=7.0cm]{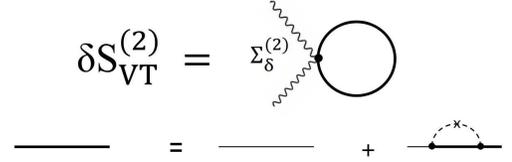}}
  \caption{Diagrammatic formalism for $\delta{S}^{(2)}_{\rm VT}$ (i.e., contribution directly from $\Sigma_{\delta}^{(2)}$ vertex). In the figure, the thin and thick solid lines denote the bare and renormalized Green function, respectively; the wavy line and dashed line with cross are associated with the nonequilibrium self-energy 
  $\Sigma_{\delta}^{(2)}$ and impurity interaction, respectively. }    
\label{figyw2}
\end{figure}

The nonequilibrium part in the effective action reads 
\begin{eqnarray}
\delta{S}&=&\!\!-\int{dx}\Big[\sum_{n=2}^{\infty}\frac{1}{n}{\rm{\bar Tr}}\Big([G_0(\Sigma_{\delta}\!+\!V\tau_3)]^n-(G_0V\tau_3)^n\Big)\nonumber\\
  &&\mbox{}\!\!+\!{\rm{\bar Tr}}(G_0\Sigma_{\delta})\!+\!\eta_f\Sigma_{\delta3}\!+\!\frac{2\Delta_0\delta|\Delta|+\delta|\Delta|^2}{g}\Big].~~~~  \label{FFaction}
\end{eqnarray}
 In principle, one only needs to consider the linear response of the weak nonequilibrium variation, i.e., keep up to the second order of the weak nonequilibrium variation in the nonequilibrium action. Then, by expanding the nonequilibrium self-energy as $\Sigma_{\delta}=\Sigma_{\delta}^{(1)}+\Sigma_{\delta}^{(2)}$ with $\Sigma_{\delta}^{(1)}$ and $\Sigma_{\delta}^{(2)}$ denoting the parts from the linear and second orders of the variation, respectively, the nonequilibrium action in Eq.~(\ref{FFaction}) becomes
\begin{equation}
\delta{S}^{(2)}=\delta{S}^{(2)}_{\rm CR}+\delta{S}^{(2)}_{\rm VT}\!-\!\int\!{dx}\Big(\eta_f\Sigma^{(2)}_{\delta3}+\frac{\delta|\Delta|^2}{g}\Big),  \label{Faction} 
\end{equation}
with the contribution of the $\Sigma_{\delta}^{(1)}$-$\Sigma_{\delta}^{(1)}$ correlation:
\begin{eqnarray}
  &&\!\!\!\!\!\!\delta{S}^{(2)}_{\rm CR}\!=\!-\frac{1}{2}\!\int\!\!{dx}{\rm {\bar Tr}}\Big[\Sigma_{\delta}^{(1)}G_0\Sigma_{\delta}^{(1)}G_0\!+\!2(G_0V\tau_3)^2(G_0\Sigma_{\delta}^{(1)})^2\!\nonumber\\
  &&\!\!\!\!\!\!\mbox{}+\!G_0\Sigma_{\delta}^{(1)}G_0V\tau_3G_0\Sigma_{\delta}^{(1)}G_0V\tau_3+O\big(V^{n>2}\big)\Big], \label{Sd1}
\end{eqnarray}
as well as a direct $\Sigma_{\delta}^{(2)}$-vertex contribution:
\begin{equation}
  \delta{S}^{(2)}_{\rm VT}\!=\!-\!\!\int\!{dx}{\rm{\bar Tr}}[(G_0\Sigma^{(2)}_{\delta})\!+\!G_0\Sigma^{(2)}_{\delta}(G_0V\tau_3)^2\!+\!O(V^{n>2})].  \label{Sd2}
\end{equation}
Consequently, from the nonequilibrium action $\delta{S}^{(2)}$, by determining the corresponding nonequilibrium self-energy, one can derive the property of the nonequilibrium variation as well as the related scattering effect. Specifically, the $\Sigma_{\delta}^{(1)}$-$\Sigma_{\delta}^{(1)}$ correlation $\delta{S}^{(2)}_{\rm CR}$ in Eq.~(\ref{Sd1}) is illustrated in Fig.~\ref{figyw1} by a connected Feynman diagram of the correlation. Corresponding to Fig.~\ref{figyw1}, on the right-hand side of Eq.~(\ref{Sd1}), the first term denotes the bare $\Sigma_{\delta}^{(1)}$-$\Sigma_{\delta}^{(1)}$ correlation; the second and third terms represent the Born and vertex corrections by impurity scattering\cite{MP}, respectively. Whereas the $\Sigma_{\delta}^{(2)}$-vertex contribution $\delta{S}^{(2)}_{\rm VT}$ in Eq.~(\ref{Sd2})  is illustrated by the Feynman diagram in Fig.~\ref{figyw2} with a renormalized bubble. As seen from the figure, the impurity interaction in $\delta{S}^{(2)}_{\rm VT}$ makes no contribution to the non-equilibrium property, and only provides the renormalization to the fermion bubble, which is same as the one by the equilibrium impurity self-energy.

\section{Transport Behavior}

In this section, we focus on the transport behavior in the diamagnetic response of superconductors. Physically, the current is conventionally expressed as ${\bf j}=\langle\psi|{\bm \Pi}|\psi\rangle/m$, with the kinematical momentum operator ${\hat {\bf \Pi}}={{\bf {\hat p}}-e{\bf A}\tau_3}$.  Among this expression, the current vertex ${\bf {\hat p}}/m$ drives a current ${\bf j}_d$ through the current-current correlation within the path-integral approach\cite{EPM,aa2} or Kubo formula\cite{MP}. The $e{\bf A}\tau_3/{m}$ part is related to the density vertex $\tau_3$\cite{Ba0} and directly pumps a current ${\bf j}_p=-e^2n{\bf A}/m$, which is considered as an unphysical non-gauge-invariant current in the literature\cite{G1,MP}. In normal metals, at the stationary case, the drive current ${\bf j}_d$ 
exactly cancels the pump current ${\bf j}_p$, and hence, the total current vanishes as it should be, since the stationary magnetic vector potential can not drive the normal-state current. Whereas in superconductors, only a part of ${\bf j}_d$ cancels ${\bf j}_p$\cite{EPM}, and then, the diamagnetic superfluid current ${\bf j}=-{en_s{\bf A}}/{m}$ emerges in the remaining part of the drive current ${\bf j}_d$.  

For the convenience of analysis and understanding, we first derive the equilibrium supercurrent ${\bf j}=en_s{\bf p}_s/m$ and hence superfluid density $n_s$ from the equilibrium action $S_0$ [Eq.~(\ref{S0})] at clean case. After that, with impurities, we discuss the nonequilibrium transport behavior by considering a nonequilibrium variation $\delta{\bf p}_s$ generated by $e\delta{\bf A}$.  It is shown that in the calculation based on kinematical momentum operator $\delta{\hat {\bf \Pi}}={{\bf {\hat p}}-e\delta{\bf A}\tau_3}$, the scattering treatment/correction leads to a mismatch in the cancellation process of the non-gauge-invariant pump current, because of the non-commutative $[{\bf {\hat p}},V({\bf r})\tau_3]$ and commutative $[e\delta{\bf {A}}\tau_3,V({\bf r})\tau_3]$. As a consequence, an unphysical current emerges. To fix this issue, we apply a special unitary transformation to eliminate the non-gauge-invariant current vertex $e\delta{\bf A}\tau_3/m$, and obtain the Meissner supercurrent vertex. Then, by performing an analytical calculation of the supercurrent-supercurrent correlation with Born and vertex corrections from the impurity scattering, the microscopic momentum-relaxation rate of superfluid is obtained to compare with GIKE and explain the relaxation-time approximation in the previous diagrammatic formulations\cite{MB,MBo,Ba1} as well as the friction resistance of Meissner supercurrent revealed in the previous works\cite{MB,MBo,Ba1,G1,Silaev,Eilen1,Eilen,Usadel}. It is further proved that the applied unitary transformation is actually equivalent to the Wilson-line technique\cite{Wilson} for the diamagnetic response.

\subsection{Equilibrium transport property}
\label{TPE}

From the equilibrium action in Eq.~(\ref{S0}) at clean case, the supercurrent is given by
\begin{eqnarray}
  {\bf j}=-e\partial_{{\bf p}_s}S_0={\bf j}_d+{\bf j}_p, 
\end{eqnarray}
where  
\begin{eqnarray}
  {\bf j}_p&=&-\frac{2e{\bf p}_s}{m}\frac{\partial{S_0}}{\partial({{p}_s^2/m})}\nonumber\\
&=&\!\sum_{p_n,{\bf k}}\frac{2e{\bf p}_s(\xi_{\bf k}\!+\!p_s^2/2m)}{m(ip_n\!-\!E_{\bf k}^+)(ip_n\!-\!E_{\bf k}^-)}\!+\!\frac{\eta_fe{\bf p}_s}{m}\!=\!\frac{en{\bf p}_s}{m}, ~~~~\label{jp}
\end{eqnarray}
and
\begin{eqnarray}
  {\bf j}_d\!\!&=&-e{\bf v_k}\frac{\partial{S_0}}{\partial({\bf v_k}\cdot{\bf p}_s)}=\sum_{p_n,{\bf k}}\frac{2e{\bf v_k}(ip_n\!-\!{\bf v_k}\!\cdot\!{\bf p}_s)}{(ip_n\!-\!E_{\bf k}^+)(ip_n\!-\!E_{\bf k}^-)}\nonumber\\
  \!\!&=&\!\!\!\frac{2ek_F^2{\bf p}_s}{3m^2}\!\!\sum_{\bf k}\partial_{E_{\bf k}}(E_{\bf k}F_{\bf k})\nonumber\\
\!\!&=&\!\!\frac{2ek_F^2{\bf p}_s}{3m^2}\sum_{\bf k}\Big[\frac{\Delta_0^2}{E_{\bf k}}\partial_{E_{\bf k}}F_{\bf k}+\partial_{\xi_{\bf k}}(\xi_{\bf k}F_{\bf k})\Big]\nonumber\\
\!\!&=&\!\!\frac{en_s{\bf p}_s}{m}-\frac{en{\bf p}_s}{m}. \label{jd}
\end{eqnarray}
Here, the superfluid density $n_s$ is given by
\begin{equation}
n_s=\frac{2k_F^2}{3m}\sum_{\bf k}\frac{\Delta_0^2}{E_{\bf k}}\partial_{E_{\bf k}}F_{\bf k},   
\end{equation}
which is exactly same as the results obtained in the previous works\cite{G1,GIKE1,GIKE5,aa2} by various approaches.

It is noted that ${\bf j}_d$ is associated with the drive current mentioned above, since it arises from the second order of the current-vertex-related term ${\bf p}_s\cdot{{\bf v_k}}$. Whereas ${\bf j}_p$ comes from the density-vertex-related term ${\bf p}_s^2\tau_3/(2m)$ and corresponds to the non-gauge-invariant pump current. Then, it is clearly seen that in normal metals with the vanishing superfluid density ($n_s=0$),  the drive current ${\bf j}_d$ 
exactly cancels the pump current ${\bf j}_p$, and hence, the total current vanishes. Whereas in superconductors, only the second-term in ${\bf j}_d$ [Eq.~(\ref{jd})] cancels ${\bf j}_p$\cite{EPM}, and then, the superfluid current ${\bf j}=-{en_s{\bf A}}/{m}$ emerges in the remaining part [first term in Eq.~(\ref{jd})] of ${\bf j}_d$.

\subsection{Issue of gauge-invariance breaking in the conventional current-current correlation}
\label{sec-issue}

We next discuss the nonequilibrium property. Based on the Hamiltonian in Eq.~(\ref{BdG}), considering a variation of the superconducting momentum, i.e., ${\bf p}_s\rightarrow{\bf p}_s+\delta{\bf p}_s(x)$, the nonequilibrium self-energy reads 
\begin{equation}\label{Issue}
  \Sigma_{\delta}=\delta{\bf p}_s\cdot{\bf v_{\hat p}}+\delta{\bf p}_s\cdot{\bf v}_{{\bf p}_s}\tau_3+\frac{{\delta}p_s^2\tau_3}{2m}\approx\delta{\bf p}_s\cdot{\bf v_{\hat p}}+\frac{{\delta}p_s^2\tau_3}{2m},
\end{equation}
in which we have neglected the term $\delta{\bf p}_s\cdot{\bf v}_{{\bf p}_s}\tau_3$ in comparison to $\delta{\bf p}_s\cdot{\bf v_{\hat p}}$ since ${p_s}\ll{k_F}$ in conventional superconductors. 

One then has the current-vertex-related term $\Sigma_{\delta}^{(1)}=\delta{\bf p}_s\cdot{\bf v_{\hat p}}$ and density-vertex-related one $\Sigma_{\delta}^{(2)}={{\delta}p_s^2\tau_3}/{(2m)}$, which therefore contribute to the drive $\delta{\bf j}_d$ and pump $\delta{\bf j}_p$ currents through the corresponding current-current correlation in $S^{(2)}_{\rm CR}$ [Eq.~(\ref{Sd1})/Fig.~\ref{figyw1}] and density-vertex contribution in $S^{(2)}_{\rm VT}$ [Eq.~(\ref{Sd2})/Fig.~\ref{figyw2}], respectively. Nevertheless, as pointed out in Sec.~\ref{sec-df}, the correlation contribution $S^{(2)}_{\rm CR}$ experiences the Born and vertex corrections by impurity scattering, and hence the drive current $\delta{\bf j}_d$ experiences the scattering influence. Whereas the impurity interaction in the vertex contribution $S^{(2)}_{\rm VT}$ makes no contribution to the non-equilibrium property, except for the normalization to the corresponding vertex. For density vertex in this circumstance, the renormalization on charge density has been revealed to vanish in the literature\cite{ISE1,ISE2,ISE3,ISE4}. Therefore, one has the pump current $\delta{\bf j}_p=en{\delta{\bf p}_s}/m$ free from the scattering influence. 

As mentioned above, the non-gauge-invariant pump current $\delta{\bf j}_{p}$ needs to be canceled by the corresponding charge-density part in the drive current $\delta{\bf j}_{d}$, so that only the contribution of the superfluid density retains in the diamagnetic response. Nevertheless, with impurities, $\delta{\bf j}_{d}$ experiences the scattering influence but $\delta{\bf j}_{p}$ does not, directly leading to a mismatch in the cancellation process. This mismatch arises from the breaking of the gauge invariance by scattering treatment. Specifically, in the diamagnetic response, the prerequisite for $\delta{\bf j}_{p}$ to exactly cancel the corresponding charge-density part in $\delta{\bf j}_{d}$ requires a gauge-invariant expected value $\langle\psi|{\bf {\hat p}}-e\delta{\bf A}\tau_3|\psi\rangle/m$ of current. But due to the non-commutative $[{\bf {\hat p}},V({\bf r})\tau_3]$ and commutative $[e\delta{\bf {A}}\tau_3,V({\bf r})\tau_3]$, the scattering treatment only plays a role in $\langle\psi|{\bf {\hat p}}|\psi\rangle$ but makes zero influence on $\langle\psi|e\delta{\bf {A}}\tau_3|\psi\rangle$, leading to the gauge-invariance breaking of the expected value of current.

To solve this issue, Abrikosov and Gorkov applied an approximation\cite{G1} that assumes an isotropic Green function at dirty case in order to first integrate over the momentum variable as mentioned in the introduction. Then, one can distinguish the contributions from superfluid density $n_s$ and total charge density $n$ in $\delta{\bf j}_{d}$, and eliminate the scattering effect in the later contribution to cancel the non-gauge-invariant pump current $\delta{\bf j}_{p}$. Whereas the transport-equation formalism\cite{EG,GIKE1} applies the Wilson-line\cite{Wilson} technique. This technique by constructing the gauge-invariant basis $|\psi_g\rangle$ leads to a gauge-invariant current vertex ${\hat {\bf j}}_g$, and the non-gauge-invariant current part $e\delta{\bf {A}}\tau_3/m$ naturally vanishes. 

\subsection{Supercurrent-supercurrent correlation}
\label{sec-ssc}

In this part, we apply a special unitary transformation to eliminate the non-gauge-invariant current vertex $e\delta{\bf A}\tau_3/m$, and obtain the Meissner supercurrent vertex. Then, one can calculate the supercurrent-supercurrent correlation with Born and vertex corrections from the impurity scattering, and obtain the microscopic momentum-relaxation rate of superfluid. 

Specifically, from Eq.~(\ref{SC}), considering a variation $\delta{\bf p}_s$ of the superconducting momentum, the action is written as
\begin{eqnarray}
  \!S&\!=\!&\int\!{dx}\Big\{\sum_{s=\uparrow,\downarrow}\!\!\psi^*_s(x)[i\partial_{x_0}\!-\!\xi_{\hat {\bf p}+{\bf p}_s+\delta{\bf p}_s}\!-\!V(x)]\psi_s(x)\nonumber\\
  &&\mbox{}-\!\psi^{\dagger}(x)\Delta_0\tau_1\psi(x)\!-\!\frac{\Delta_0^2}{g}\Big\}.
\end{eqnarray}
Applying the unitary transformation 
\begin{equation}\label{ut}
  \psi(x)\rightarrow\exp\Big[i\tau_3\int^{\bf x}_0{\delta{\bf p}_s({\bf x}')}{\cdot}d{\bf x'}\Big]\psi(x),  
\end{equation}
the action becomes
\begin{widetext}
\begin{eqnarray}
  S\!&=&\!\!\int\!{dx}\psi_g^{\dagger}(x)\Big[i\partial_{x_0}\!-\!\xi_{\hat {\bf p}+{\bf p}_s\tau_3}\tau_3\!-\!V(x)\tau_3\!-\!\Delta_0\tau_+\exp\Big(2i\int^{{\bf x}}_0\delta{\bf p}_s{\cdot}d{\bf x'}\Big)\!-\!\Delta_0\tau_-\exp\Big(\!-\!2i\int^{{\bf x}}_0\delta{\bf p}_s{\cdot}d{\bf x'}\Big)\Big]\psi_g(x)\nonumber\\
  &&\mbox{}\!\!-\!\!\int\!{dx}\Big[\frac{\Delta_0^2}{g}+\eta_fV(x)\Big]. \label{a1}
\end{eqnarray}
\end{widetext}
Here, we only focus on the stationary diamagnetic response and neglect the electric-field effect.

Consequently, from the action in Eq.~(\ref{a1}), for the small variation, one finds the non-equilibrium self-energy:
\begin{equation}\label{sss}
  \Sigma_{\delta}=-2\Delta_0\tau_2\int^{\bf x}_0\delta{\bf p}_s\cdot{\bf dx'}.
\end{equation}
Then, in comparison to Eq.~(\ref{Issue}) based on the conventional current vertex, the density-vertex-related term $\delta{p}^2_s\tau_3/m$ that is related to the non-gauge-invariant pump current vanishes in Eq.~(\ref{sss}), and hence, there is no vertex contribution $\delta{S}^{(2)}_{\rm VT}$. Particularly, it is noted that at long-wave limit, the derived self-energy in Eq.~(\ref{sss}) becomes
\begin{equation}\label{SVT}
\Sigma_{\delta}^{(1)}\!=\!-2\Delta_0\tau_2\delta{\bf p}_s\!\cdot\!{\hat {\bf x}}=2i\Delta_0\tau_2\delta{\bf p}_s\!\cdot\!\partial_{\hat {\bf k}},
\end{equation}
which describes the drive effect by $\delta{\bf p}_s$ in the diamagnetic response to generate the Meissner supercurrent\cite{G1}. We therefore refer to $i2\Delta_0\tau_2\partial_{\hat {\bf k}}$ as the Meissner-supercurrent vertex. 

Consequently, with the Meissner-supercurrent vertex [Eq.~(\ref{SVT})],  one can derive the supercurrent-supercurrent correlation $\delta{S}^{(2)}_{\rm CR}$ [Eq.~(\ref{Sd1})/Fig.~\ref{figyw1}] with the Born and vertex corrections by impurity scattering. By assuming an adiabatic variation $\delta{\bf p}_s(x)=\delta{\bf p}_se^{iqz+0^+x_0}$ with $z$ being the spatial coordinate along the direction perpendicular to the surface, at the weak impurity interaction, after the summation of the Matsubara frequency, the current $\delta{\bf j}=-e\partial_{\delta{\bf p}_s^*(x)}\delta{S^{(2)}}=-e\partial_{\delta{\bf p}_s}\delta{S^{(2)}}/2$ is derived as (refer to Appendix~\ref{a-ssc})
\begin{eqnarray}
&&\!\!\!\!\delta{\bf j}\!=\!\sum_{\bf k}{e{\bf v_k}}\Big\{\rho_{{\bf k}}\!+\frac{n_i\pi}{i{\zeta}_{\bf k}v_Fq}\!\sum_{\bf k'}|V_{\bf kk'}|^2(\rho_{{\bf k}}\!-\!\rho_{{\bf k'}})\nonumber\\
    &&\mbox{}\!\!\!\!\times\Big[\sum_{\eta=\pm}\Big(e^-_{\bf kk'}\delta(E_{\bf k}^{\eta}-E_{\bf k'}^{\eta})\!+\!e^+_{\bf kk'}\delta(E_{\bf k}^{\eta}\!-\!E_{\bf k'}^{-\eta})\Big)\Big]\Big\},~~~\label{finalcurrent}
\end{eqnarray}
with $\rho_{{\bf k}}=({{\bf v_k}\cdot\delta{\bf p}_s})\frac{2\Delta_0^2}{E_{\bf k}}\partial_{E_{\bf k}}F_{\bf k}$ and $e^{\pm}_{\bf kk'}=\frac{1}{2}(1\pm\frac{\Delta_0^2}{E_{\bf k}E_{\bf k'}})$; $\zeta_{\bf k}$ represents a coefficient (refer to Appendix~\ref{a-ssc}). It is noted that $\zeta_{\bf k}{v}_F{q}$ is a diffusive pole, which emerges at the stationary diffusion case. The first and second terms on the right-hand side of Eq.~(\ref{finalcurrent}) represent the source and scattering terms, respectively, both of which exactly recover the ones from GIKE\cite{GIKE1,GIKE5}.

Further considering the fact that {\small $(\rho_{{\bf k}}\!-\!\rho_{{\bf k}'})[\delta(E^{+}_{\bf k}\!-\!E^{+}_{\bf k'})\!+\!\delta(E^{-}_{\bf k}\!-\!E^{-}_{\bf k'})]$}
in the scattering term of Eq.~(\ref{finalcurrent}) vanishes around the Fermi surface, the current becomes
\begin{eqnarray}
\delta{\bf j}\!&=&\!\sum_{\bf k}{e{\bf v_k}}\Big\{\rho_{{\bf k}}\!+\!\frac{n_i\pi}{i{\zeta}_{\bf k}v_Fq}\sum_{\bf k'}|V_{\bf kk'}|^2(\rho_{{\bf k}}\!-\!\rho_{{\bf k'}})\nonumber\\
    &&\mbox{}\!\times\Big[\sum_{\eta=\pm}e^+_{\bf kk'}\delta(E_{\bf k}^{\eta}\!-\!E_{\bf k'}^{-\eta})\Big]\Big\}. \label{jj}
\end{eqnarray}

As pointed out in Refs.~\onlinecite{GIKE1,GIKE5}, the scattering term in above equation is finite only at the emergence of the normal fluid, which requires $|{\bf v_k}\cdot{\bf p}_s|>E_{\bf k}$ as mentioned in Sec.~\ref{secHA}. Whereas this condition requires a threshold $p_L=\Delta_0/v_F$ for the superconducting momentum $p_s$ to exceed. Therefore, at $p_s<\Delta_0/v_F$, one has
\begin{equation}
\delta{\bf j}=\frac{en_s\delta{\bf p}_s}{m},  
\end{equation}
which is free from the diffusive influence by impurity scattering, showing the superconductivity phenomenon.

Whereas for $p_s>\Delta_0/v_F$, as pointed out in Ref.~~\onlinecite{GIKE1,GIKE5}, the current is captured by three-fluid (normal fluid as well as viscous and non-viscous superfluids) model and can be divided into three parts:
\begin{equation}
\delta{\bf j}=\delta{\bf j}_{vs}+\delta{\bf j}_{nvs}+\delta{\bf j}_n,  
\end{equation}
with
\begin{eqnarray}
  \delta{\bf j}_{vs}&=&\!\sum_{{\bf k}\in{\rm P_{v}}}{e{\bf v_k}}\rho_{{\bf k}}\Big(1+\frac{\Gamma_{\bf k}}{i{\zeta}_{\bf k}v_Fq}\Big),\label{jvs}\\
  \delta{\bf j}_{nvs}&=&\!\sum_{{\bf k}\in{\rm P_{nv}}}{e{\bf v_k}}\rho_{{\bf k}},\label{jnvs}\\
   \delta{\bf j}_n&=&\!-\sum_{{\bf k}\in{\rm U}}\frac{e{\bf v_k}}{i{\zeta}_{\bf k}v_Fq}\sum_{{\bf k'}\in{\rm P_v}}\rho_{\bf k'}{D_{\bf kk'}},\label{jn}
\end{eqnarray}
and
\begin{eqnarray}\label{TK}
  \Gamma_{\bf k}&=&n_i\pi\sum_{{\bf k'}\in{U}}|V_{\bf kk'}|^2e^+_{\bf kk'}\Big[\sum_{\eta=\pm}\delta(E_{\bf k}^{\eta}\!-\!E_{\bf k'}^{-\eta})\Big],\\
  D_{\bf kk'}&=&{n_i\pi}|V_{\bf kk'}|^2\Big[\sum_{\eta=\pm}e^+_{\bf kk'}\delta(E_{\bf k}^{\eta}\!-\!E_{\bf k'}^{-\eta})\Big]. ~~~~\label{GK}
\end{eqnarray}
Here, $n_i$ denotes the impurity density and $N(0)$ represents the density of states; $\Gamma_{\bf k}$ stands for the microscopic momentum-relaxation rate of superfluid; $D_{\bf kk'}$ represents the microscopic friction rate between superfluid and normal fluid;   ${\rm P_{nv}}$ denotes the non-viscous pairing regions in momentum space with finite $F_{\bf k}$ but zero $\Gamma_{\bf k}$; ${\rm P_v}$ represents the viscous pairing regions with both finite $F_{\bf k}$ and $\Gamma_{\bf k}$;  $U$ stands for the unpairing region with vanishing $F_{\bf k}$.

Specifically, in the source term on the right-hand side of Eq.~(\ref{jj}),  only particles in the pairing regions with nonzero anomalous correlation $F_{\bf k}$ are driven by $\delta{\bf p}_s$ to contribute to the current. For ${\bf k}$ particle lying in the pairing region, one has $E_{\bf k}^+>0$ and $E_{\bf k}^-<0$ as mentioned in Sec.~\ref{secHA}. In this circumstance, in the scattering term, once the energy conservation can not be satisfied for any ${\bf k}'$, the ${\bf k}$ particle is free from the momentum-relaxation scattering, and one therefore gets the non-viscous superfluid and hence the current ${\bf j}_{nvs}$ in Eq.~(\ref{jnvs}). But once the energy conservation is satisfied to give rise to nonzero scattering term, one finds $E_{\bf k'}^->0$ by $\delta(E_{\bf k}^+-E_{\bf k'}^-)$ or $E_{\bf k'}^+<0$ by $\delta(E_{\bf k}^--E_{\bf k'}^+)$, and hence, ${\bf k'}$ particle lies in the unpairing region (normal fluid) with vanishing $F_{\bf k'}$ and hence $\rho_{{\bf k'}}$. This scattering between particles in pairing and unpairing regions, behaves like the friction between superfluid and normal fluid, leading to the viscous superfluid and the current ${\bf j}_{vs}$ in Eq.~(\ref{jvs})

It is noted that for particles in the unpairing regions (normal fluid), although the source term in Eq.~(\ref{jj}) is zero as it should be, but due to the friction mentioned above, the scattering term is finite. Specifically, if ${\bf k}$ particle lies in the unpairing region with $E_{\bf k}^->0$ or $E_{\bf k}^+<0$, according to the energy conservation in the scattering term, ${\bf k'}$ particle can lie in both viscous pairing and unpairing regions, i.e., the particles from normal fluid experience the scattering from those in both viscous superfluid and normal fluid. The scattering between particles in normal fluid is natural but makes zero contribution to current as $\rho_{{{\bf k}\in{U}}}=\rho_{{\bf k'}\in{U}}=0$ in the scattering term. But through the friction drag with the viscous superfluid current, a normal-fluid current is induced in Eq.~(\ref{jn}).

Based on the analysis above, the total current at $p_s>\Delta_0/v_F$ can be re-written as
\begin{equation}
\delta{\bf j}=\frac{en^{\rm eff}_s\delta{\bf p}_s}{m}+\frac{\Gamma^{\rm eff}_c}{iv_Fq}\frac{en^{\rm eff}_s\delta{\bf p}_s}{m},  \label{sj}
\end{equation}
with the effective superfluid density:
\begin{equation}
n_s=\sum_{{\bf k}\in{{\rm P_{nv}}+{\rm P_{v}}}}\frac{2k_F^2\cos\theta_{\bf k}^2}{m}\frac{\Delta_0^2}{E_{\bf k}}\partial_{E_{\bf k}}F_{\bf k},   
\end{equation}
and effective current-relaxation rate:
\begin{equation}
\Gamma^{\rm eff}_c=\frac{\sum_{{\bf k}\in{\rm P_{v}}}{\bf v_{\bf k}}\rho_{{\bf k}}\Gamma_{\bf k}/\zeta_{\bf k}-\sum_{{\bf k}\in{\rm U},{\bf k'}\in{\rm P_v}}{{\bf v_k}}\rho_{{\bf k'}}{D_{\bf kk'}/\zeta_{\bf k}}}{\sum_{{\bf k}\in{{\rm P_{nv}}+{\rm P_{v}}}}{{\bf v_k}}\rho_{{\bf k}}}.
\end{equation}
We then obtain the equation of motion of the supercurrent with the influence of the scattering. Then, in real space, Eq.~(\ref{sj}) becomes a diffusive equation:
\begin{equation}
\partial_z\delta{\bf j}=\partial_z\delta{\bf p}_s\frac{en^{\rm eff}_s}{m}+\frac{\Gamma^{\rm eff}_c}{{v_F}}\frac{en^{\rm eff}_s\delta{\bf p}_s}{m},  \label{dej}
\end{equation}
which in consideration of the fact $\partial_z\delta{\bf p}_s=\partial_z{\bf p}_s$ is equivalent to:
\begin{equation}
\partial_z^2{\bf j}=\partial_z^2{\bf p}_s\frac{en^{\rm eff}_s}{m}+\frac{\Gamma^{\rm eff}_c}{{v_F}}\frac{en^{\rm eff}_s\partial_z{\bf p}_s}{m},  \label{dejj}
\end{equation}
Consequently, we arrives at a self-consistent equation of motion of the superconducting momentum/current with the influence of the scattering. Clearly, the second term on the right-hand side of above equation denotes the friction resistance of the supercurrent. As mentioned above, this resistance [$\Gamma_{\bf k}$ in Eq.~(\ref{TK}) and $D_{\bf k}$ in Eq.~(\ref{GK})] is nonzero only with the emergence of the normal fluid at $p_s>\Delta_0/v_F$.  Therefore, as pointed out in Ref.~\onlinecite{GIKE1}, the friction resistance of the Meissner supercurrent in the diamagnetic response emerges only when the superconducting velocity is larger than a threshold at which the normal fluid emerges, similar to the Landau's theory for the emerged fluid viscosity in bosonic liquid helium at larger velocity\cite{Landau}. 

Together with the Maxwell equation, the penetration depth from Eq.~(\ref{dejj}) is derived as 
\begin{equation}
\lambda=\lambda_c\sqrt{\frac{1}{1-\frac{\xi}{l}}}\approx\lambda_c\sqrt{{1+\frac{\xi}{l}}},  
\end{equation}
where the coherence length $\xi=\lambda/\kappa$ as well as mean free path $l=v_F/(\kappa\Gamma^{\rm eff}_c)$ and clean-limit penetration depth {${\lambda_c}=\sqrt{m/(4{\pi}e^2n^{\rm eff}_s)}$}, with $\kappa$ being the Ginzburg-Landau parameter. Then, the previously revealed phenomenological dependence of the penetration depth on mean-free path by Tinkham\cite{Ba1} is recovered within the diagrammatic formalism at weak scattering case, same as the formulation within the GIKE\cite{GIKE1}. 

{\sl Role of Doppler shift.}---It is noted that the Doppler shift plays two important roles in the derivation/results above. On one hand, as mentioned in Sec.~\ref{secHA}, it leads to the generation of the normal fluid at $p_s>\Delta_0/v_F$. On the other hand, it guarantees the vanishing intraband scattering in Eq.~(\ref{finalcurrent}) by $(\rho_{\bf k}-\rho_{\bf k'})\delta(E_{\bf k}^{\eta}-E_{\bf k'}^{\eta})$ part around Fermi surface, and hence, only the interband scattering by $(\rho_{\bf k}-\rho_{\bf k'})\delta(E_{\bf k}^{\eta}-E_{\bf k'}^{-\eta})$ part retains. This interband scattering occurs only when $p_s$ is larger than the threshold $\Delta_0/v_F$\cite{GIKE1}, and hence, the momentum relaxation of superfluid current emerges only at $p_s>\Delta_0/v_F$ (where the normal fluid emerges). However, in the previous formulation of the scattering in superconductors\cite{G1}, the Doppler shift was approximately neglected. As a consequence, the normal fluid dynamics is absent. Most importantly, in this circumstance, the interband scattering in Eq.~(\ref{finalcurrent}) by $(\rho_{\bf k}-\rho_{\bf k'})\delta(E_{\bf k}^{\eta}-E_{\bf k'}^{-\eta})=(\rho_{\bf k}-\rho_{\bf k'})\delta(\sqrt{\Delta_0^2+\xi_k^2}+\sqrt{\Delta_0^2+\xi_{k'}^2})$ part is forbidden, but the interband scattering by $(\rho_{\bf k}-\rho_{\bf k'})\delta(E_{\bf k}^{\eta}-E_{\bf k'}^{\eta})=(\rho_{\bf k}-\rho_{\bf k'})\delta(\sqrt{\Delta_0^2+\xi_k^2}-\sqrt{\Delta_0^2+\xi_{k'}^2})$ part is always finite around Fermi surface. Consequently, the superfluid current in the diamagnetic response always experiences the friction resistance by impurity scattering as the theoretical descriptions in the previous works\cite{G1,Silaev,Eilen1,Eilen,Usadel} revealed, in contrast to the superconductivity phenomenon.

\subsection{A gauge-invariant description}

In the previous part, within the diagrammatic formalism, by applying the unitary transformation in Eq.~(\ref{ut}), the non-gauge-invariant-current vertex $e{\bf A}\tau_3/m$, i.e., the issue of gauge-invariance breaking by scattering treatment as mentioned in Sec.~\ref{sec-issue}, is eliminated. To understand this unitary transformation, following the treatment within the transport-equation formalism\cite{EG,GIKE1}, we next apply the Wilson-line\cite{Wilson} technique to construct the gauge-invariant field operator and provide a gauge-invariant description to handle the nonequilibrium transport property in superconductors.

We begin with the action in consideration of a four-vector variation $e\delta{A}_{\mu}=(e\delta\phi,e\delta{\bf A})$ of electromagnetic potential:
\begin{eqnarray}\label{action1}
  \!S&\!=\!&\!\!\!\int\!{dx}\!\!\!\sum_{s=\uparrow,\downarrow}\!\!\psi^*_s(x)[i\partial_{x_0}\!-\!\xi_{\hat {\bf p}+{\bf p}_s-e\delta{\bf A}}\!-\!e\delta{\phi}(x)]\psi_s(x)\nonumber\\
  &&\mbox{}-\!\int{dx}\Big[\psi^{\dagger}(x){\hat \Delta}(x)\psi(x)\!+\!\frac{|\Delta(x)|^2}{g}\Big]\nonumber\\
    &\!=\!&\!\!\!\int\!{dx}\Big[\psi^{\dagger}(x){\hat G^{-1}}\psi(x)\!-\!\eta_f\frac{({\bf p}_s\!-\!e\delta{\bf A})^2}{2m}\!-\!\frac{|\Delta(x)|^2}{g}\Big].~~~~~~
\end{eqnarray}
Here, ${\hat G^{-1}}=i\partial_{x_0}-\xi_{\hat {\bf p}+{\bf p}_s\tau_3-e\delta{\bf A}\tau_3}\tau_3\!-\!e\delta{\phi}(x)\tau_3-{\hat \Delta}(x)$; ${\hat \Delta}(x)={\Delta}(x)\tau_++{\Delta}^*(x)\tau_-$, where the superconducting order parameter ${\Delta}(x)=[\Delta_0+\delta|\Delta|(x)]e^{i\delta\theta(x)}$ with $\Delta_0$ and $\delta|\Delta|(x)$ as well as $\delta\theta(x)$ denoting the equilibrium gap and nonequilibrium Higgs mode\cite{Am0,OD1,OD2,OD3,Am12,Am3,Am4,Am5,Am6,symmetry} as well as the superconducting phase fluctuation\cite{Am0,gi0,AK,Gm1,Gm2,Ba0,pm0,pi1,pi2,gi1,AHM}, respectively. In addition, we expand the scalar potential as
\begin{equation}
  \delta\phi(x)=\delta\phi_0(x_0)+\int^{\bf x}_{0}\nabla_{\bf x'}\delta\phi(x_0,{\bf x'})d{\bf x'},
\end{equation}
in order to distinguish the Josephson voltage effect\cite{Josephson} by $\delta\phi_0(x_0)$ and electric-field drive effect by $\nabla_{\bf x}\delta\phi(x_0,{\bf x})$.

It is noted that under a gauge transformation:
\begin{equation}\label{gaugestructure0}
  \psi(x){\rightarrow}e^{i\tau_3\chi(x)}\psi(x),
\end{equation}  
the action in Eq.~(\ref{action1}) satisfies the gauge structure in superconductors first revealed by Nambu\cite{gi0,gi1}:
\begin{eqnarray}
e{\delta}A_{\mu}&\rightarrow&e{\delta}A_{\mu}-\partial_{\mu}\chi(x), \label{gaugestructure1}\\
\label{gaugestructure2}
\delta\theta(x)&\rightarrow&\delta\theta(x)+2\chi(x),
\end{eqnarray}
where the four-vector $\partial_{\mu}=(\partial_{x_0},-{\bm \nabla})$.

Under the gauge transformations in Eqs.~(\ref{gaugestructure0})-(\ref{gaugestructure2}), the conventional Wilson-line\cite{Wilson} technique to construct gauge-invariant field operator $\psi_g=e^{i\tau_3P\int^{x}_0d{x}^{\mu}e{\delta{A}_{\mu}}}\psi$ is difficult to handle for deriving the gauge-invariant Kernel ${\hat G^{-1}}$ and performing the further calculation within the diagrammatic formalism. To simplify the formulation,  we restrict the gauge-transformation function $\chi(x)$ to depend on either spatial coordinate or time coordinate. Then, one can apply a simplified Wilson-line technique to construct the gauge-invariant field operator:
\begin{equation}
  \psi_g(x)\!=\!\exp\Big[i\tau_3e\Big(\!\int^{x_0}_0\!\!\!\delta\phi_0dx'_0-\!\!\int^{{\bf x}}_0\!\!\!\delta{\bf A}{\cdot}d{\bf x'}\Big)\Big]\psi(x).  \label{WLT}
\end{equation}
Consequently, on basis of the gauge-invariant $\psi_g(x)$, the action in Eq.~(\ref{action1}) becomes
\begin{equation}
  S\!=\!\!\int\!{dx}\Big[\psi_g^{\dagger}(x){\hat G}^{-1}_g(x)\psi_g(x)-\frac{|\Delta(x)|^2}{g}\Big],
\end{equation}
with the Green-function Kernel:
\begin{widetext}
\begin{eqnarray}\label{gi-action}
{\hat G}^{-1}_g(x)&=&i\partial_{x_0}\!-\!\xi_{\hat {\bf p}+{\bf p}_s\tau_3}\tau_3-\int_0^{\bf x}e{\bf E}{\cdot}d{\bf x'}\!-\!|\Delta|\tau_+\exp\Big(i\delta\theta+2ie\!\int^{x_0}_0\delta\phi_0dx'_0-2ie\int^{{\bf x}}_0\delta{\bf A}{\cdot}d{\bf x'}\Big)\nonumber\\
  &&\mbox{}-|\Delta|\tau_-\exp\Big(-i\delta\theta-2ie\!\int^{x_0}_0\delta\phi_0dx'_0+2ie\int^{{\bf x}}_0\delta{\bf A}{\cdot}d{\bf x'}\Big). 
\end{eqnarray}
\end{widetext}
Here, ${\bf E}=-\nabla_{\bf x}\delta\phi-\partial_{x_0}\delta{\bf A}$ denotes the gauge-invariant electric field. It is noted that ${\hat G}^{-1}_g(x)$ is directly gauge invariant under the gauge transformations in Eqs.~(\ref{gaugestructure1}) and (\ref{gaugestructure2}). Particularly, in the derived ${\hat G}^{-1}_g(x)$ via Wilson-line technique, there is no non-gauge-invariant-current (density-vertex-related) term, similar to the derivation within the transport-equation formalism\cite{GIKE1}. Whereas in diamagnetic response, the Wilson line technique in Eq.~(\ref{WLT}) reduces to the unitary transformation in Eq.~(\ref{ut}), and hence, the derivation applying this unitary transformation in Sec.~\ref{sec-ssc} avoids the issue of gauge-invariance breaking by scattering treatment mentioned in Sec.~\ref{sec-issue}.  

Furthermore, it is established the superconductors can directly respond to vector potential ${\bf A}$ (Meissner effect/Ginzburg-Landau kinetic term) in addition to the electric field ${\bf E}=-{\nabla}_{\bf R}\phi-\partial_t{\bf A}$, differing from normal metals that solely respond to electric field. Whereas the conventional calculation with the vector potential alone is hard to distinguish these two effects in superconductors. Therefore, ${\hat G}^{-1}_g(x)$ in Eq.~(\ref{gi-action}) provides an efficient Lagrangian/Hamiltonian Kernel, which explicitly distinguish the drive effect by the electric field $e{\bf E}$ and the Meissner effect driven by effective vector potential\cite{G1}:
\begin{equation}
i\delta\theta\!-\!2ie\!\int^{{\bf x}}_0\!\delta{\bf A}{\cdot}d{\bf x'}=2i\!\int^{{\bf x}}_0\!\big(\nabla_{\bf x}\delta\theta/2\!-\!e\delta{\bf A}\big)d{\bf x'},  
\end{equation}
as well as the Josephson effect induced by effective electric voltage\cite{Josephson}:
\begin{equation}
i\delta\theta+2ie\!\int^{x_0}_0\delta\phi_0dx'_0=2i\int^{x_0}_0(\partial_{x_0}\delta\theta/2+e\delta\phi_0), 
\end{equation}
and all these characteristic effects manifest themselves in a gauge-invariant description. One therefore expects a wide application of this Kernel to study the mesoscopic physics in superconductors as well as more diagrammatic-formalism and transport-equation investigations.

\section{Higgs mode}

We next focus on the Higgs mode. Based on the BCS Hamiltonian in Eq.~(\ref{BdG}), considering a variation of the superconducting gap (i.e., Higgs mode), the nonequilibrium self-energy is derived as $\Sigma_{\delta}(x)=\delta|\Delta|(x)\tau_1$. In this circumstance, $\delta{S^{(2)}_{\rm CR}}$ in Eq.~(\ref{Sd1}) denotes the contribution from the amplitude-amplitude correlation with the Born and vertex corrections by impurity scattering. 

For free case, we take the superconducting momentum ${\bf p}_s=0$.  Then, in center-of-mass frequency-momentum space [$x=(x_0,{\bf x})\rightarrow{q}=(\Omega,{\bf q})$], at weak impurity interaction and long-wave limit (${\bf q}=0$), after the summation of the Matsubara frequency, one has (refer to Appendix~\ref{a-aac})
\begin{eqnarray}\label{HNS}
  \delta{S}^{(2)}&=&-\int{d\Omega}|\delta\Delta|^2\Big\{[(2\Delta_0)^2-\Omega^2]\sum_{\bf k}\frac{\partial_{E_{\bf k}}F_{\bf k}}{4E_{\bf k}}-i\Omega{n_i}\pi\nonumber\\
  &&\mbox{}\times\sum_{{\bf kk'}}|V_{\bf kk'}|^2\frac{\Delta_0^2\xi_{\bf k}^2}{4E_{\bf k}^6}F_{\bf k}\delta(E_{\bf k}-E_{\bf k'})\Big\}.~~~~  
\end{eqnarray}
Here, $F_{\bf k}=[f(E_{\bf k})-f(-E_{\bf k})]/(2E_{\bf k})$.  It is pointed out that on the right-hand side of above equation, the scattering part (second term) arises from the vertex correction solely, whereas the Born correction makes no contribution at all. This is because that the Born correction is equivalent to the renormalization of the equilibrium impurity self-energy and hence vanishes according to the Anderson theorem\cite{ISE,ISE1,ISE2,ISE3,ISE4}. Whereas the vertex correction that only emerges at nonequilibrium case breaks the time-translational symmetry, as mentioned in the introduction, and hence, makes a finite contribution to the Higgs-mode damping.

Furthermore, from $\partial_{\delta|\Delta|}\delta{S^{(2)}}=0$, the equation of motion of the Higgs mode is given by 
\begin{equation}\label{EOMHM}
\big[(2\Delta_0)^2-\Omega^2-2i\Omega\gamma_H\big]\delta|\Delta|=0,  
\end{equation}
with the Higgs-mode damping rate:
\begin{equation}
\gamma_H\approx\frac{\Gamma_0\sum_{\bf k}{\Delta_0^2|\xi_{\bf k}|F_{\bf k}}/{E_{\bf k}^5}}{\sum_{\bf k}4\partial_{E_{\bf k}}F_{\bf k}/E_{\bf k}}.  
\end{equation}
Here, $\Gamma_0=2n_i{\pi}D\int\frac{d\Omega_{\bf k'}}{4\pi}|V_{\bf k_F-k_F'}|^2$. It is noted that Eq.~(\ref{EOMHM}) at clean limit reduces to the previously revealed one by various theoretical approaches in the literature\cite{Silaev,aa2,GIKE2,GIKE4,GIKE3,Cea1,Cea2,aa1,Cea3,Am0,Am12,EPM,Am3,Am4,Am5,Am6,symmetry}, showing a gapful energy spectrum $\omega_H=2\Delta_0$. With impurities, the emerged term $2i\Omega\gamma_H$ in Eq.~(\ref{EOMHM}), which is proportional to $\Omega$, suggests that the impurity scattering effect on Higgs mode is a nonequilibrium property with the time-translational-symmetry breaking. The derived damping rate $\gamma_H$ is exactly same as the one obtained from GIKE\cite{GIKE3}, and as mentioned in Ref.~\onlinecite{GIKE3}, due to this damping, the long-time dynamic of the Higgs mode after excitation behaves as
\begin{equation}
\delta|\Delta|(t)\sim\frac{\cos(2\Delta_0t)e^{-{\bar \gamma}_Ht}}{\sqrt{\Delta_0t}},  
\end{equation}
where ${\bar \gamma}_H$ is the average of $\gamma_H$ in the momentum space. Therefore, in contrast to the coherent BCS oscillatory decay ${\cos(2\Delta_0t)}/{\sqrt{\Delta_0t}}$ at clean limit, the impurity scattering leads to the faster exponential decay.

The induced damping of the Higgs mode by impurity scattering agrees with the analysis through Heisenberg equation of motion as mentioned in the introduction, since the  Higgs-mode excitation and electron-impurity interaction are non-commutative in Nambu space. Whereas as mentioned in the introduction, the previous derivation of the Higgs mode within the Eilenberger equation\cite{Silaev} fails to give this damping and derives a Higgs-mode energy spectrum that is free from the scattering influence. Actually, this is because that the microscopic scattering integral in Eilenberger equation is incomplete. As proved in Ref.~\onlinecite{EF},  because of the quasiclassical approximation on $\tau_3$-Green function, the scattering integral in Eilenberger equation only involves the anisotropic part of the Green function that is related to the transport property, but generically drops out the isotropic one which determines the Higgs mode\cite{EF}.

\section{Summary}

In summary, we have analytically performed a diagrammatic formulation of the impurity scattering in superconductors, as a separate but complementary approach to compare with GIKE\cite{GIKE1,GIKE3}. Both transport and collective Higgs mode are addressed, in order to fill the gap in the textbook calculation of the Kubo current-current correlation in superconductors with the impurity scattering, and resolve the controversy (whether the impurity scattering can lead to the damping of Higgs mode) between GIKE\cite{GIKE3} and Eilenberger equation\cite{Silaev} in the literature.

For transport behavior in the diamagnetic response,  within the conventional calculation based on kinematical momentum operator ${\hat {\bf \Pi}}={{\bf {\hat p}}-e{\bf A}\tau_3}$,  it is shown that a non-gauge-invariant current emerges after the scattering treatment/correction in the current-current correlation.
In order to resolve this issue of the gauge-invariance breaking, we apply a special unitary transformation that is equivalent to the Wilson-line technique for diamagnetic response, and obtain the Meissner-supercurrent vertex. Then, the supercurrent-supercurrent correlation with the Born and vertex corrections from impurity scattering is formulated. Particularly, in contrast to the previous works\cite{G1} in the literature that overlooked the Doppler shift, we keep this effect in the quasiparticle energy spectra. Then, the previously revealed microscopic momentum-relaxation rate of superfluid and the current captured by three-fluid (normal fluid as well as viscous and non-viscous superfluids) model\cite{GIKE1,GIKE5} are exactly recovered. The momentum-relaxation rate of superfluid is finite only when the superconducting momentum is larger than a threshold $\Delta_0/v_F$, at which the normal fluid emerges and causes the friction with the superfluid current, similar to the Landau's superfluid theory of bosonic liquid helium\cite{Landau}. This derivation uncovers the physics behind the relaxation-time approximation in the previous diagrammatic formulations\cite{MB,MBo,Ba1}, which leads to the friction resistance of the Meissner supercurrent. It is also pointed out that the Doppler shift is essential to guarantee the vanishing
momentum-relaxation rate of superfluid at small superconducting velocity. Whereas in the previous theoretical descriptions\cite{G1,Silaev,Eilen1,Eilen,Usadel} that overlooked this effect, the derived superfluid current always experiences the friction resistance by impurity scattering as a consequence, in contrast to the superconductivity phenomenon. Furthermore, through the Wilson-line technique, a gauge-invariant Hamiltonian that explicitly distinguishes the Meissner effect and electric-field drive effect as well as the Josephson voltage effect  is proposed.   

As for the collective Higgs mode, we calculate the amplitude-amplitude correlation with the Born and vertex corrections from impurity scattering. The vertex correction, which only emerges at nonequilibrium case with time-translational-symmetry breaking, leads to a fast exponential damping of the Higgs mode, whereas the Born correction that is equivalent to equilibrium impurity self-energy makes no contribution because of the Anderson theorem\cite{ISE,ISE1,ISE2,ISE3,ISE4}.  The derived damping by impurity scattering from the diagrammatic formalism exactly recovers the one from GIKE\cite{GIKE3} and agrees with the analysis through Heisenberg equation of motion, but is in contrast to the vanishing one obtained in Eilenberger equation\cite{Silaev}. The reason leading to missing damping is due to the generically incomplete scattering integral in Eilenberger equation\cite{EF}. The life-time of Higgs mode due to the impurity scattering provides a possible origin for the experimentally observed broadening of the resonance signal\cite{NL3,NL4} as well as the damping after optical excitation\cite{NL1,NL2} of the Higgs mode. Moreover, as pointed out in Ref.~\onlinecite{GIKE3}, the damping by impurities can cause a phase shift in the optical signal of Higgs mode, which exhibits a $\pi$-jump at the resonance frequency and hence provides a very clear feature for further experimental detection.

\begin{acknowledgments}
The authors acknowledge financial support from
the National Natural Science Foundation of 
China under Grants No.\ 11334014 and No.\ 61411136001.  
\end{acknowledgments}

\begin{widetext}
\begin{appendix}

\section{Derivation of charge density}
\label{a-cd}

In this part, we present the derivation of the charge density. With the density vertex $\tau_3$ in Nambu space\cite{Ba0}, in the effective nonequilibrium action in Eq.~(\ref{Faction}), the contribution from the density-vertex-related part of the nonequilibrium self-energy reads $\{{\rm {\bar Tr}}[G_0(p)\tau_3]+\eta_f\}\Sigma_{\delta3}$. In this contribution, substituting the Green function in Eq.~(\ref{Greenfunction}), one finds the prefactor: 
\begin{equation}
{\rm {\bar Tr}}[G_0(p)\tau_3]+\eta_f=\sum_{p_n,{\bf k}}\frac{2(\xi_{\bf k}\!+\!p_s^2/2m)}{m(ip_n\!-\!E_{\bf k}^+)(ip_n\!-\!E_{\bf k}^-)}\!+\!\eta_f=\sum_{\bf k}\Big[1\!+\!2\Big(\xi_{\bf k}\!+\!\frac{p_s^2}{2m}\Big)F_{\bf k}\Big]=-\frac{2k_F^2}{3m}\sum_{{\bf k}}\partial_{\xi_{\bf k}}(\xi_{\bf k}F_{\bf k})\approx\frac{2k_F^2N(0)}{3m},
\end{equation}
which is exactly the charge density $n$.

\section{Derivation of supercurrent-supercurrent correlation}
\label{a-ssc}

In this part, we derive the supercurrent-supercurrent correlation with Born and vertex corrections from the impurity scattering. Specifically, for transport behavior in the diamagnetic response, substituting the derived self-energy in Eq.~(\ref{SVT}) that is related to the the Meissner-supercurrent vertex,  the supercurrent-supercurrent correlation $S^{(2)}_{\rm CR}$ [Eq.~(\ref{Sd1})/Fig.~\ref{figyw1}] is written as
\begin{eqnarray}
  \delta{S}^{(2)}_{\rm CR}&=&2\Delta_0^2\int{dx}{\rm {\bar Tr}}\Big[\tau_2(\delta{\bf p}_s\!\cdot\!{\bf v}_{\hat {\bf k}})\partial_{\xi_{\bf k}}G_0\tau_2(\delta{\bf p}_s\!\cdot\!{\bf v}_{\hat {\bf k}})\partial_{\xi_{\bf k}}G_0\!+\!2(G_0V\tau_3)^2(G_0\tau_2\delta{\bf p}_s\!\cdot\!{\bf v}_{\hat {\bf k}}\partial_{\xi_{\bf k}})^2\!\nonumber\\
    &&\mbox{}+\!G_0\tau_2(\delta{\bf p}_s\!\cdot\!{\bf v}_{\hat {\bf k}})\partial_{\xi_{\bf k}}G_0V\tau_3G_0\tau_2(\delta{\bf p}_s\!\cdot\!{\bf v}_{\hat {\bf k}})\partial_{\xi_{\bf k}}G_0V\tau_3\Big]=I_{\rm ba}+I_{\rm bc}+I_{\rm vc},\label{SCR}
\end{eqnarray}
with the bare supercurrent-supercurrent correlation $I_{\rm ba}$ as well as Born $I_{\rm bc}$ and vertex $I_{\rm vc}$ corrections written as
\begin{eqnarray}
  I_{\rm ba}\!\!\!&=&\!\!{2\Delta_0^2}\!\!\!\sum_{ip_n,{\bf k}}\Big[(\delta{\bf p}_s\!\cdot\!{\bf v}_{{\bf k}})^2{\rm Tr}[\tau_2\partial_{\xi_{\bf k}}G_0(ip_n^+,{\bf k}^+)\tau_2\partial_{\xi_{\bf k}}G_0(ip_n,{\bf k})],\label{ba00}\\
    I_{\rm bc}\!\!\!&=&\!\!-{2\Delta_0^2}\!\!\!\sum_{ip_n,{\bf kk'}}(\delta{\bf p}_s\!\cdot\!{\bf v}_{{\bf k}})^2n_i|V_{\bf kk'}|^2\Big\{{\rm Tr}[\partial_{\xi_{\bf k}}G_0(ip_n,{\bf k})\tau_2G_0(ip_n^+,{\bf k}^+)\tau_2\partial_{\xi_{\bf k}}G_0(ip_n,{\bf k})\tau_3G(ip_n,{\bf k'})\tau_3]\!+\!(p^+\rightarrow{p^-})\Big\},~~~~~~\label{bc0}\\
    I_{\rm vc}\!\!\!&=&\!\!-{2\Delta_0^2}\!\!\!\sum_{ip_n,{\bf kk'}}(\delta{\bf p}_s\!\cdot\!{\bf v}_{{\bf k}})(\delta{\bf p}_s\!\cdot\!{\bf v}_{{\bf k'}})n_i|V_{\bf kk'}|^2{\rm Tr}[\partial_{\xi_{\bf k}}G_0(ip_n,{\bf k})\tau_2G_0(ip_n^+,{\bf k}^+)\tau_3G_0(ip_n^+,{\bf k'}^+)\tau_2\partial_{\xi_{\bf k'}}G_0(ip_n,{\bf k'})\tau_3].\label{vc0}
\end{eqnarray}
Here, we have kept up to the second order of the impurity interaction by considering the case of weak impurity scattering. Here, $p^{\pm}=(ip_n^{\pm},{\bf k}^{\pm})=(ip_n{\pm}i0^+,{\bf k}\pm{\bf q})$.

With the Green function in Eq.~(\ref{Greenfunction}), around the Fermi surface, one has $\partial_{\xi_{\bf k}}G_0(ip_n,{\bf k})\approx{\tau_3}/{\Lambda_{\bf k}(ip_n)}$ with $\Lambda_{\bf k}(ip_n)=(ip_n-E_{\bf k}^+)(ip_n-E_{\bf k}^-)$. Then, after the summation of Matsubara frequency, the bare supercurrent-supercurrent correlation [Eq.~(\ref{ba00})] reads
\begin{eqnarray}\label{fba}
  I_{\rm ba}=-\sum_{ip_n,{\bf k}}\frac{{2\Delta_0^2}2(\delta{\bf p}_s\!\cdot\!{\bf v}_{{\bf k}})^2}{\Lambda_{\bf k}(ip_n)\Lambda_{\bf k^+}(ip_n^+)}\approx-{2\Delta_0^2}\sum_{ip_n,{\bf k}}\frac{2(\delta{\bf p}_s\!\cdot\!{\bf v}_{{\bf k}})^2}{\Lambda^2_{\bf k}(ip_n)}=-\sum_{\bf k}(\delta{\bf p}_s\!\cdot\!{\bf v}_{{\bf k}})^2\frac{2\Delta_0^2}{E_{\bf k}}\partial_{E_{\bf k}}F_{\bf k}.
\end{eqnarray}

Moreover, using the fact $\frac{1}{\Lambda_{\bf k}(ip_n)}=\frac{1}{2E_{\bf k}}\sum_{\eta=\pm}\frac{\eta}{ip_n-E_{\bf k}^{\eta}}$, the Born [Eq.~(\ref{bc0})] and vertex [Eq.~(\ref{vc0})] corrections by impurity scattering are given by  
\begin{eqnarray}
\!\!\!\!I_{\rm bc}&=&-{4\Delta_0^2}\sum_{ip_n,{\bf kk'}}(\delta{\bf p}_s\!\cdot\!{\bf v}_{{\bf k}})^2n_i|V_{\bf kk'}|^2\Big[\frac{(ip_n^+-{\bf v_{k^+}}\!\cdot\!{\bf p}_s)(ip_n-{\bf v_{k'}}\!\cdot\!{\bf p}_s)-\Delta_0^2}{\Lambda^2_{\bf k}(ip_n)\Lambda_{\bf k^+}(ip_n^+)\Lambda_{\bf k'}(ip_n)}+(p^+\rightarrow{p^-})\Big]\nonumber\\
  &=&{4\Delta_0^2}\sum_{ip_n,{\bf kk'},\eta\lambda}(\delta{\bf p}_s\!\cdot\!{\bf v}_{{\bf k}})^2n_i|V_{\bf kk'}|^2\Big[\frac{\Delta_0^2\!-\!(ip_n^+\!-\!{\bf v_{k^+}}\!\cdot\!{\bf p}_s)(ip_n\!-\!{\bf v_{k'}}\!\cdot\!{\bf p}_s)}{4E_{\bf k'}E_{\bf k^+}\Lambda^2_{\bf k}(ip_n)}\frac{\eta}{ip_n^+\!-\!E_{\bf k^+}^{\eta}}\frac{\lambda}{ip_n\!-\!E_{\bf k'}^{\lambda}}\!+\!(p^+\rightarrow{p^-})\Big],\label{bc1}\\
 I_{\rm vc}&=&{4\Delta_0^2}\sum_{ip_n,{\bf kk'},\eta\lambda}(\delta{\bf p}_s\!\cdot\!{\bf v}_{{\bf k}})(\delta{\bf p}_s\!\cdot\!{\bf v}_{{\bf k'}})n_i|V_{\bf kk'}|^2\frac{(ip_n^+-{\bf v_{k^+}}\!\cdot\!{\bf p}_s)(ip_n^+-{\bf v_{k'^+}}\!\cdot\!{\bf p}_s)-\Delta_0^2}{\Lambda_{\bf k^+}(ip_n^+)\Lambda_{\bf k'}(ip_n)\Lambda_{\bf k}(ip_n)\Lambda_{\bf k'^+}(ip_n^+)}\nonumber\\
 &=&{4\Delta_0^2}\sum_{ip_n,{\bf kk'}}(\delta{\bf p}_s\!\cdot\!{\bf v}_{{\bf k}})(\delta{\bf p}_s\!\cdot\!{\bf v}_{{\bf k'}})n_i|V_{\bf kk'}|^2\frac{(ip_n^+-{\bf v_{k^+}}\!\cdot\!{\bf p}_s)(ip_n^+-{\bf v_{k'^+}}\!\cdot\!{\bf p}_s)-\Delta_0^2}{4E_{\bf k^+}E_{\bf k'^+}\Lambda_{\bf k'}(ip_n)\Lambda_{\bf k}(ip_n)}\frac{\eta}{ip_n^+-E_{\bf k^+}^{\eta}}\frac{\lambda}{ip_n^+-E_{\bf {k'}^+}^{\lambda}}.\label{vc1}
\end{eqnarray}

Further considering the imaginary (i.e., scattering) parts of the Born and vertex corrections, through the summation of the Matsubara frequency, one has   
\begin{eqnarray}
\!\!\!\!I_{\rm bc}&=&-i\pi{4\Delta_0^2}\sum_{ip_n,{\bf kk'},\eta'\eta\lambda}(\delta{\bf p}_s\!\cdot\!{\bf v}_{{\bf k}})^2n_i|V_{\bf kk'}|^2\eta'\eta\lambda\Big[\frac{\Delta_0^2-(ip_n-{\bf v_{k^+}}\!\cdot\!{\bf p}_s)(ip_n-{\bf v_{k'}}\!\cdot\!{\bf p}_s)}{4E_{\bf k'}E_{\bf k^+}2E_{\bf k}\Lambda_{\bf k}(ip_n)(ip_n-E_{\bf k}^{\eta'})}\frac{\delta(ip_n\!-\!E_{\bf k^+}^{\eta})}{ip_n\!-\!E_{\bf k'}^{\lambda}}\!-\!({\bf k^+}\!\!\rightarrow\!{\bf k^-})\Big]\nonumber\\
  &=&-i\pi{4\Delta_0^2}\sum_{{\bf kk'},\eta\eta'\lambda}(\delta{\bf p}_s\!\cdot\!{\bf v}_{{\bf k}})^2n_i|V_{\bf kk'}|^2\eta\eta'\lambda\Big[\frac{\Delta_0^2\!-\!(E_{\bf k'}^{\lambda}\!-\!{\bf v_{k^+}}\!\cdot\!{\bf p}_s){\lambda}E_{\bf k'}}{4E_{\bf k'}E_{\bf k^+}\Lambda_{\bf k}(E_{\bf k'}^{\lambda})(E_{\bf k'}^{\lambda}\!-\!E_{\bf k}^{\eta'})}\frac{f(E_{\bf k'}^{\lambda}){\delta(E_{\bf k'}^{\lambda}\!-\!E_{\bf k^+}^{\eta})}}{2E_{\bf k}}\!-\!({\bf k^+}\!\!\rightarrow\!{\bf k^-})\Big]\nonumber\\
&=&i\pi{4\Delta_0^2}\sum_{{\bf kk'},\eta\eta'\lambda}(\delta{\bf p}_s\!\cdot\!{\bf v}_{{\bf k}})^2\frac{n_i|V_{\bf kk'}|^2}{4E_{\bf k}}\eta'\Big[e^{-\eta\lambda}_{\bf k^+k'}\frac{f(E_{\bf k^+}^{\eta}){\delta(E_{\bf k'}^{\lambda}\!-\!E_{\bf k^+}^{\eta})}}{\Lambda_{\bf k}(E_{\bf k^+}^{\eta})(E_{\bf k^+}^{\eta}\!-\!E_{\bf k}^{\eta'})}\!-\!({\bf k^+}\!\rightarrow\!{\bf k^-})\Big],
  \label{bc2}
\end{eqnarray}
and 
\begin{eqnarray}
  I_{\rm vc}&=&i\pi{4\Delta_0^2}\!\!\!\!\sum_{\eta\lambda,ip_n{\bf kk'}}\!\!\eta\lambda(\delta{\bf p}_s\!\cdot\!{\bf v}_{{\bf k}})(\delta{\bf p}_s\!\cdot\!{\bf v}_{{\bf k'}})n_i|V_{\bf kk'}|^2\frac{\Delta_0^2\!-\!(ip_n\!-\!{\bf v_{k^+}}\!\cdot\!{\bf p}_s)(ip_n\!-\!{\bf v_{k'^+}}\!\cdot\!{\bf p}_s)}{4E_{\bf k^+}E_{\bf k'^+}\Lambda_{\bf k'}(ip_n)\Lambda_{\bf k}(ip_n)}\Big[\frac{\delta(ip_n\!-\!E_{\bf k^+}^{\eta})}{ip_n\!-\!E_{\bf {k'}^+}^{\lambda}}\!+\!\frac{\delta(ip_n\!-\!E_{\bf {k'}^+}^{\lambda})}{ip_n\!-\!E_{\bf k^+}^{\eta}}\Big]\nonumber\\
  &=&i\pi{4\Delta_0^2}\sum_{ip_n,{\bf kk'},\eta\lambda}\eta\lambda(\delta{\bf p}_s\!\cdot\!{\bf v}_{{\bf k}})(\delta{\bf p}_s\!\cdot\!{\bf v}_{{\bf k'}})n_i|V_{\bf kk'}|^2\frac{\Delta_0^2\!-\!(ip_n\!-\!{\bf v_{k^+}}\!\cdot\!{\bf p}_s)(ip_n\!-\!{\bf v_{k'^+}}\!\cdot\!{\bf p}_s)}{4E_{\bf k^+}E_{\bf k'^+}\Lambda_{\bf k'}(ip_n)\Lambda_{\bf k}(ip_n)}\frac{2\delta(ip_n\!-\!E_{\bf k^+}^{\eta})}{ip_n\!-\!E_{\bf {k'}^+}^{\lambda}}\nonumber\\
  &=&i\pi{8\Delta_0^2}\sum_{ip_n,{\bf kk'},\eta\lambda}(\delta{\bf p}_s\!\cdot\!{\bf v}_{{\bf k}})(\delta{\bf p}_s\!\cdot\!{\bf v}_{{\bf k'}})n_i|V_{\bf kk'}|^2\frac{\Delta_0^2\!-\!(ip_n\!-\!{\bf v_{k^+}}\!\cdot\!{\bf p}_s)(ip_n\!-\!{\bf v_{k'^+}}\!\cdot\!{\bf p}_s)}{4E_{\bf k^+}E_{\bf k'^+}2E_{\bf k'}\Lambda_{\bf k}(ip_n)}\frac{\eta\lambda\lambda'\delta(ip_n\!-\!E_{\bf k^+}^{\eta})}{(ip_n\!-\!E_{\bf {k'}^+}^{\lambda})(ip_n\!-\!E_{\bf {k'}}^{\lambda'})}\nonumber\\
  &=&-i\pi{8\Delta_0^2}\sum_{{\bf kk'},\eta\lambda\lambda'}\frac{(\delta{\bf p}_s\!\cdot\!{\bf v}_{{\bf k}})(\delta{\bf p}_s\!\cdot\!{\bf v}_{{\bf k'}})n_i|V_{\bf kk'}|^2}{\Lambda_{\bf k}(E_{\bf k^+}^{\eta})}\Big[\frac{\lambda'e_{\bf k^+k'^+}^{-\eta\lambda}f(E_{\bf {k'}^+}^{\lambda})\delta(E_{\bf {k'}^+}^{\lambda}\!-\!E_{\bf k^+}^{\eta})}{4E_{\bf k'}(E_{\bf {k'}^+}^{\lambda}\!-\!E_{\bf {k'}}^{\lambda'})}\!-\!\frac{{\lambda}e_{\bf k^+k'^+}^{-\eta\lambda'}f(E_{\bf {k'}}^{\lambda'})\delta(E_{\bf {k'}}^{\lambda'}\!-\!E_{\bf k^+}^{\eta})}{4E_{\bf k'}(E_{\bf {k'}^+}^{\lambda}\!-\!E_{\bf {k'}}^{\lambda'})}\Big]\nonumber\\
   &=&-i\pi{8\Delta_0^2}\sum_{{\bf kk'},\eta\lambda\lambda'}\frac{(\delta{\bf p}_s\!\cdot\!{\bf v}_{{\bf k}})(\delta{\bf p}_s\!\cdot\!{\bf v}_{{\bf k'}})n_i|V_{\bf kk'}|^2}{\Lambda_{\bf k}(E_{\bf k^+}^{\eta})}e_{\bf k^+k'^+}^{-\eta\lambda}\lambda'\Big[\frac{f(E_{\bf {k'}^+}^{\lambda})\delta(E_{\bf {k'}^+}^{\lambda}\!-\!E_{\bf k^+}^{\eta})}{4E_{\bf k'}(E_{\bf {k'}^+}^{\lambda}\!-\!E_{\bf {k'}}^{\lambda'})}\!-\!\frac{f(E_{\bf {k'}}^{\lambda})\delta(E_{\bf {k'}}^{\lambda}\!-\!E_{\bf k^+}^{\eta})}{4E_{\bf k'}(E_{\bf {k'}^+}^{\lambda'}\!-\!E_{\bf {k'}}^{\lambda})}\Big].
\label{vc2}
\end{eqnarray}
in which we have used the fact $\delta(E_{\bf k^+}^{\eta'}-E_{\bf k}^{\eta})\equiv0$.

For long-wave case, the leading contribution in Eq.~(\ref{bc2}) comes from $\eta=\eta'$ part, and the Born correction becomes 
\begin{eqnarray}
  I_{\rm bc}&\approx&i\pi{4\Delta_0^2}\sum_{{\bf kk'},\eta\lambda}(\delta{\bf p}_s\!\cdot\!{\bf v}_{{\bf k}})^2\frac{n_i|V_{\bf kk'}|^2}{4E_{\bf k}}\eta\Big[\frac{e_{\bf kk'}^{-\eta\lambda}f(E_{\bf k^+}^{\eta})}{(E_{\bf k^+}^{\eta}\!-\!E_{\bf k}^{\eta})^2(E_{\bf k^+}^{\eta}\!-\!E_{\bf k}^{-\eta})}\!-\!({\bf k^+}\!\rightarrow\!{\bf k^-})\Big]{\delta(E_{\bf k'}^{\lambda}\!-\!E_{\bf k}^{\eta})} \nonumber\\
   &\approx&i\pi{4\Delta_0^2}\sum_{\eta\lambda}\sum_{{\bf kk'}}(\delta{\bf p}_s\!\cdot\!{\bf v}_{{\bf k}})^2\frac{n_i|V_{\bf kk'}|^2}{4E_{\bf k}}\eta\Big[\frac{e_{\bf kk'}^{-\eta\lambda}f(E_{\bf k^+}^{\eta})}{(E_{\bf k^+}^{\eta}\!-\!E_{\bf k}^{\eta})^2}\Big(\frac{1}{2\eta{E_{\bf k}}}\!-\!\frac{E_{\bf k^+}^{\eta}\!-\!E_{\bf k}^{\eta}}{4E_{\bf k}^2}\Big)\!-\!({\bf k^+}\!\rightarrow\!{\bf k^-})\Big]{\delta(E_{\bf k'}^{\lambda}\!-\!E_{\bf k}^{\eta})}
  \nonumber\\
  &=&i\pi{8\Delta_0^2}\sum_{\eta\lambda}\sum_{{\bf kk'}}(\delta{\bf p}_s\!\cdot\!{\bf v}_{{\bf k}})^2\frac{n_i|V_{\bf kk'}|^2}{4E_{\bf k}}\eta\Big[\frac{e_{\bf kk'}^{-\eta\lambda}f(E_{\bf k^+}^{\eta})}{(E_{\bf k^+}^{\eta}\!-\!E_{\bf k}^{\eta})^2}\Big(\frac{1}{2\eta{E_{\bf k}}}\!-\!\frac{E_{\bf k^+}^{\eta}\!-\!E_{\bf k}^{\eta}}{4E_{\bf k}^2}\Big)\!-\!({\bf k^+}\!\rightarrow\!{\bf k})\Big]{\delta(E_{\bf k'}^{\lambda}\!-\!E_{\bf k}^{\eta})}\nonumber\\
  &\approx&-\pi\sum_{\eta\lambda{\bf kk'}}(\delta{\bf p}_s\!\cdot\!{\bf v}_{{\bf k}})^2{n_i|V_{\bf kk'}|^2e_{\bf kk'}^{-\eta\lambda}}\frac{2\Delta_0^2}{E_{\bf k}}\partial_{E_{\bf k}}\Big[\frac{{\eta}f(E_{\bf k}^{\eta})}{E_{\bf k}}\Big]\frac{\delta(E_{\bf k'}^{\lambda}\!-\!E_{\bf k}^{\eta})}{i\zeta_{\bf k}{v_Fq}}. \label{fbc}
\end{eqnarray}
Whereas in Eq.~(\ref{vc2}), both $\lambda'=\lambda$ and $\lambda'=-\lambda$ parts play an important role, and the vertex correction reads
\begin{eqnarray}
  I_{\rm vc}&\approx&-i\pi{8\Delta_0^2}\sum_{{\bf kk'},\eta\lambda}\frac{(\delta{\bf p}_s\!\cdot\!{\bf v}_{{\bf k}})(\delta{\bf p}_s\!\cdot\!{\bf v}_{{\bf k'}})n_i|V_{\bf kk'}|^2}{(E_{\bf k^+}^{\eta}\!-\!E_{\bf k}^{\eta})(E_{\bf k^+}^{\eta}\!-\!E_{\bf k}^{-\eta})}e_{\bf kk'}^{-\eta\lambda}\Big[\lambda\frac{e_{\bf kk'}^{\eta\lambda}}{4E_{\bf k'}}\frac{f(E_{\bf {k'}^+}^{\lambda})\!-\!f(E_{\bf {k'}}^{\lambda})}{E_{\bf {k'}^+}^{\lambda}\!-\!E^{\lambda}_{\bf {k'}}}\!-\!\frac{f(E_{\bf {k'}^+}^{\lambda})}{4E^2_{\bf k'}}\Big]\delta(E_{\bf {k'}^+}^{\lambda}\!-\!E_{\bf k}^{\eta})\nonumber\\
  &\approx&\pi{\Delta_0^2}\sum_{{\bf kk'},\eta\lambda}{(\delta{\bf p}_s\!\cdot\!{\bf v}_{{\bf k}})(\delta{\bf p}_s\!\cdot\!{\bf v}_{{\bf k'}})n_i|V_{\bf kk'}|^2}e_{\bf kk'}^{-\eta\lambda}\frac{2\Delta_0^2}{E_{\bf k}}\partial_{E_{\bf k'}}\Big[\frac{{\eta}f(E_{\bf k'}^{\lambda})}{E_{\bf k'}}\Big]\frac{\delta(E_{\bf k'}^{\lambda}\!-\!E_{\bf k}^{\eta})}{i\zeta_{\bf k}{v_Fq}}.\label{fvc}
\end{eqnarray}
Here, $\zeta_{\bf k}{v_Fq}=2{\bf v_q}\cdot{\bf p}_s+4{\xi_{\bf k}{{\bf v_k}\cdot{\bf q}}}/E_{\bf k}+\eta({\bf v_k}\cdot{\bf q})^2/E_{\bf k}\approx2{\bf v_q}\cdot{\bf p}_s+4{\xi_{\bf k}{{\bf v_k}\cdot{\bf q}}}/E_{\bf k}+({\bf v_k}\cdot{\bf q})^2/E_{\bf k}$ at long-wave case. Consequently, with Eqs.~(\ref{fba}) and (\ref{fbc}) as well as (\ref{fvc}), the supercurrent-supercurrent correlation with Born and vertex corrections by impurity scattering and hence the current in Eq.~(\ref{finalcurrent}) are derived. 

Moreover, it is noted that ${\zeta}_{\bf k}v_Fq$ in the scattering contribution [Eqs.~(\ref{fbc}) and (\ref{fvc})] provides a diffusive pole that emerges at the stationary diffusion case. Whereas for the non-stationary case at long-wave limit [i.e., $\delta{\bf p}_s(x)=\delta{\bf p}_se^{-i(\Omega+i0^+)x_0}$], with $p^{\pm}=(ip_n^{\pm},{\bf k}^{\pm})=[ip_n{\pm}(\Omega+i0^+),{\bf k}]$ in Eq.~(\ref{SCR}), the diffusive pole ${\zeta}_{\bf k}v_Fq$ in Eqs.~(\ref{fbc}) and (\ref{fvc}) is replaced by $-\Omega$ after the derivation, and one finds a current:
\begin{equation}
\delta{\bf j}=\frac{en^{\rm eff}_s\delta{\bf p}_s}{m}-\frac{\Gamma^{\rm eff}_c}{i\Omega}\frac{en^{\rm eff}_s\delta{\bf p}_s}{m},  
\end{equation}
which is similar to the one described by Drude model.

\section{Derivation of amplitude-amplitude correlation}
\label{a-aac}

In this part, we present the derivation of the amplitude-amplitude with Born and vertex corrections from the impurity scattering. Specifically, for the collective Higgs mode, substituting the derived self-energy $\Sigma_{\delta}^{(1)}=\delta|\Delta|\tau_1$, the amplitude-amplitude correlation $S^{(2)}_{\rm CR}$ [Eq.~(\ref{Sd1})/Fig.~\ref{figyw1}] is written as 
\begin{eqnarray}
  \delta{S}^{(2)}_{\rm CR}&=&-\frac{1}{2}\int{dx}{\rm {\bar Tr}}\Big[\delta|\Delta|\tau_1G_0\delta|\Delta|\tau_1G_0\!+\!(G_0V\tau_3)^2(G_0\delta|\Delta|\tau_1)^2\!+\!G_0\delta|\Delta|\tau_1G_0V\tau_3G_0\delta|\Delta|\tau_1G_0V\tau_3\Big]\nonumber\\
  &=&-H_{\rm ba}-H_{\rm bc}-H_{\rm vc},~~~~~~\label{SCRH}
\end{eqnarray}
where the bare amplitude-amplitude correlation $H_{\rm ba}$ as well as Born $H_{\rm bc}$ and vertex $H_{\rm vc}$ corrections read
\begin{eqnarray}
  H_{\rm ba}&=&\frac{\delta|\Delta|^2}{2}\sum_{ip_n,{\bf k}}{\rm Tr}[\tau_1G_0(ip_n^+,{\bf k})\tau_1G_0(ip_n,{\bf k})]={\delta|\Delta|^2}\sum_{ip_n,{\bf k}}\frac{ip_n^+ip_n\!-\!\xi_{\bf k}^2\!+\!\Delta_0^2}{\Lambda_{\bf k}(ip_n^+)\Lambda_{\bf k}(ip_n)},\label{Hba0}\\
  H_{\rm bc}&=&\frac{\delta|\Delta|^2}{2}\sum_{ip_n,{\bf kk'}}n_i|V_{\bf kk'}|^2\big\{{\rm Tr}[\tau_1G_0(ip_n^+,{\bf k})\tau_1G_0(ip_n,{\bf k})\tau_3G(ip_n,{\bf k'})\tau_3G_0(ip_n,{\bf k})]\!+\!(p^+\rightarrow{p^-})\big\}\nonumber\\
  &=&{\delta|\Delta|^2}\sum_{ip_n,{\bf kk'}}n_i|V_{\bf kk'}|^2\Big[\frac{(ip_n^+ip_n\!-\!\xi_{\bf k}^2\!+\!\Delta_0^2)[(ip_n)^2\!-\!\Delta_0^2]\!+\!2\xi_{\bf k}^2\Delta_0^2\!+\!\Omega{ip_n}\xi_{\bf k}^2}{\Lambda_{\bf k}^2(ip_n)\Lambda_{\bf k'}(ip_n)\Lambda_{\bf k}(ip_n^+)}\!+\!(p^+\rightarrow{p^-})\Big],\label{Hbc0}\\
  H_{\rm vc}&=&\frac{\delta|\Delta|^2}{2}\sum_{ip_n,{\bf kk'}}n_i|V_{\bf kk'}|^2{\rm Tr}[G_0(ip_n,{\bf k})\tau_1G_0(ip_n^+,{\bf k}^+)\tau_3G_0(ip_n^+,{\bf k'}^+)\tau_1G_0(ip_n,{\bf k'})\tau_3]\nonumber\\
  &=&{\delta|\Delta|^2}\sum_{ip_n,{\bf kk'}}n_i|V_{\bf kk'}|^2\frac{\Delta_0^2(ip_n+ip_n^+)^2-(ip_nip_n^+-E_{\bf k}^2+2\Delta_0^2)(ip_nip_n^+-E_{\bf k'}^2+2\Delta_0^2)}{\Lambda_{\bf k}(ip_n^+)\Lambda_{\bf k'}(ip_n)\Lambda_{\bf k}(ip_n)\Lambda_{\bf k'}(ip_n^+)}.\label{Hvc0} 
\end{eqnarray}
Here, $p^{\pm}=(ip_n^{\pm},{\bf k}^{\pm})=[ip_n{\pm}(\Omega+i0^+),{\bf k}]$ and we have kept up to the second order of the impurity interaction by considering the case of weak impurity scattering. 

After the summation of Matsubara frequency, the bare amplitude-amplitude correlation
that has been well established in the literature\cite{Cea1,Cea2,aa1,Cea3,aa2,EPM} is written as
\begin{eqnarray}
  H_{\rm ba}&=&{\delta|\Delta|^2}\sum_{ip_n,{\bf k}}\frac{(ip_n^+)^2\!+\!(ip_n)^2-(ip_n^+\!-\!ip_n)^2\!-\!2E_{\bf k}^2\!+\!4\Delta_0^2}{2\Lambda_{\bf k}(ip_n^+)\Lambda_{\bf k}(ip_n)}={\delta|\Delta|^2}\sum_{ip_n,{\bf k}}\Big[\frac{4\Delta_0^2\!-\!\Omega^2}{2\Lambda_{\bf k}(ip_n^+)\Lambda_{\bf k}(ip_n)}\!+\!\frac{1}{\Lambda_{\bf k}(ip_n)}\Big]\nonumber\\
  &=&{\delta|\Delta|^2}\Big[\sum_{ip_n,{\bf k}}\frac{4\Delta_0^2\!-\!\Omega^2}{2\Lambda_{\bf k}(ip_n^+)\Lambda_{\bf k}(ip_n)}\!-\frac{1}{g}\Big]\approx{\delta|\Delta|^2}\Big[\sum_{{\bf k}}\frac{(4\Delta_0^2\!-\!\Omega^2)\partial_{E_{\bf k}}F_{\bf k}}{4E_{\bf k}}\!-\frac{1}{g}\Big],\label{Hba}
\end{eqnarray}
It is noted that the last term in above equation is canceled by the last term in Eq.~(\ref{Faction}), and then, the previously established equation of motion $[\partial_t^2-(2\Delta_0)^2]\delta|\Delta|=0$ of Higgs mode at clean case is recovered.  

We then consider the imaginary (i.e., scattering) parts of the Born [Eq.~(\ref{Hbc0})] and vertex [Eq.~(\ref{Hvc0})] corrections by using the fact $\frac{1}{\Lambda_{\bf k}(ip_n^+)}=\frac{1}{2E_{\bf k}}\sum_{\eta=\pm}\frac{\eta}{ip_{n+}+i0^+-E_{\bf k}^{\eta}}$, where $ip_{n+}=ip_n+\Omega$. 

Specifically, through the summation of the Matsubara frequency, one finds the imaginary part of the Born correction:   
\begin{eqnarray}
  H_{\rm bc}&=&-{\delta|\Delta|^2}i\pi\sum_{\eta\lambda}\eta\lambda\sum_{ip_n,{\bf kk'}}n_i|V_{\bf kk'}|^2\Big[\frac{(ip_{n+}ip_n\!-\!\xi_{\bf k}^2\!+\!\Delta_0^2)[(ip_n)^2\!-\!\Delta_0^2]\!+\!2\xi_{\bf k}^2\Delta_0^2\!+\!\Omega{ip_n}\xi_{\bf k}^2}{4E_{\bf k}E_{\bf k'}\Lambda_{\bf k}^2(ip_n)}\frac{\delta(ip_{n+}\!-\!E^{\eta}_{\bf k})}{ip_n\!-\!E^{\lambda}_{\bf k'}}\!-\!(\Omega\rightarrow{-\Omega})\Big] \nonumber\\
  &=&-{\delta|\Delta|^2}i\pi\sum_{\eta\lambda}\eta\lambda\sum_{{\bf kk'}}n_i|V_{\bf kk'}|^2\Big[\frac{(2\Delta_0^2\!-\!{\Omega}{\eta}E_{\bf k})\xi_{\bf k'}^2\!+\!2\xi_{\bf k}^2\Delta_0^2\!+\!\Omega(\eta{E_{\bf k}}\!-\!\Omega)\xi_{\bf k}^2}{4E_{\bf k}E_{\bf k'}(E_{\bf k}\!+\!E_{\bf k'})^2}\frac{f(E_{\bf k'}^{\lambda})\delta(E_{\bf k'}^{\lambda}\!+\!\Omega\!-\!E^{\eta}_{\bf k})}{(E_{\bf k}\!-\!E_{\bf k'})^2}\!-\!(\Omega\rightarrow{-\Omega})\Big] \nonumber\\
  &\approx&-{\delta|\Delta|^2}i\pi\sum_{\eta}\sum_{{\bf kk'}}n_i|V_{\bf kk'}|^2\frac{(4\Delta_0^2\!-\!{\Omega}^2)\xi_{\bf k}^2}{4E_{\bf k}E_{\bf k'}(E_{\bf k}\!+\!E_{\bf k'})^2}\Big[\frac{f(E_{\bf k'}^{\eta})\delta(E_{\bf k'}\!-\!E_{\bf k})}{\Omega^2}\!-\!\frac{f(E_{\bf k'}^{\eta})\delta(E_{\bf k'}\!-\!E_{\bf k})}{\Omega^2}\Big]=0. \label{Hbc}
\end{eqnarray}
Consequently, the Born correction to the amplitude-amplitude correlation by impurity scattering vanishes. This is because that the Born correction is equivalent to the renormalization of the equilibrium impurity self-energy. Whereas according to the Anderson theorem\cite{ISE}, this renormalization does not influence gap, and hence, makes no contribution to the damping of Higgs mode.  Actually, the bare amplitude-amplitude correlation and Born correction together can be re-written as
\begin{equation}
H_{\rm ba}+H_{\rm bc}=\frac{\delta|\Delta|^2}{2}\sum_{ip_n,{\bf k}}{\rm Tr}[\tau_1{\bar G}(ip_n^+,{\bf k})\tau_1{\bar G}(ip_n,{\bf k})], \label{RE1}
\end{equation}
with the renormalized Green function ${\bar G}=G_0+G_0V\tau_3GV\tau_3G$ (thick solid line in Fig.~\ref{figyw1}) given by\cite{ISE1,ISE2,ISE3,ISE4} ${\bar G}(ip_n,{\bf k})={[i{\bar p}_n+\xi_{\bf k}\tau_3+{\bar \Delta}_0\tau_1]}/{[(i{\bar p}_n)^2-\xi^2_{\bf k}-{\bar \Delta}_0^2]}$. Here, ${\bar p}_n$ and ${\bar \Delta}_0$ denote the renormalized Matsubara frequency and gap by impurity self-energy, respectively. It has been revealed  in the literature\cite{ISE1,ISE2,ISE3,ISE4} that ${\bar p}_n/{\bar \Delta}_0=p_n/\Delta_0$, leading to a vanishing influence from the renormalization on gap equation (Anderson theorem\cite{ISE}). Then, similar to the derivation of the bare amplitude-amplitude correlation [Eq.~(\ref{Hba})], Eq.~(\ref{RE1}) is directly derived as
\begin{eqnarray}
H_{\rm ba}+H_{\rm bc}&=&\delta|\Delta|^2\sum_{ip_n,{\bf k}}\frac{4{\bar \Delta}_0^2\!-\!{\bar \Omega}^2}{2(\xi_{\bf k}^2\!+\!{\bar \Delta}_0^2\!+\!{\bar p}_n^2)^2}=\delta|\Delta|^2\sum_{ip_n}{\pi}N(0)\frac{4{\bar \Delta}_0^2\!-\!{\bar \Omega}^2}{4({\bar \Delta}_0^2\!+\!{\bar p}_n^2)^{3/2}}=\delta|\Delta|^2(4{\Delta}_0^2\!-\!{\Omega}^2)\sum_{ip_n}\frac{{\pi}N(0){\Delta_0}/{{\bar \Delta}_0}}{4({\Delta}_0^2\!+\!{p}_n^2)^{3/2}}   \nonumber\\
&=&\delta|\Delta|^2(4{\Delta}_0^2\!-\!{\Omega}^2)\sum_{ip_n,{\bf k}}\frac{{\Delta_0}/{{\bar \Delta}_0}}{2(\xi_{\bf k}^2\!+\!{\Delta}_0^2\!+\!{p}_n^2)^2},
\end{eqnarray}
in which there is no damping term of the Higgs mode. Clearly, the Born correction makes no contribution to the damping of Higgs mode. 

The imaginary part of the vertex correction [Eq.~(\ref{Hvc0})] by impurity scattering after the summation of Matsubara frequency is written as
\begin{eqnarray}
  H_{\rm vc}&=&-{\delta|\Delta|^2}i\pi\sum_{\eta\eta'\lambda}\sum_{ip_n,{\bf kk'}}n_i|V_{\bf kk'}|^2\eta'\frac{\Delta_0^2(2{\lambda}E_{\bf k'}\!-\!\Omega)^2\!-\!(2\Delta_0^2\!-\!{\lambda}E_{\bf k'}\Omega)^2}{8E_{\bf k}^3\Lambda_{\bf k'}(E_{\bf k'}^{\lambda}\!-\!\Omega)}\frac{2\eta\lambda\delta(ip_{n+}\!-\!E_{\bf k'}^{\lambda})}{(ip_{n+}\!-\!E_{\bf k}^{\eta})(ip_n\!-\!E_{\bf k}^{\eta'})}\nonumber\\
  &=&-{\delta|\Delta|^2}i\pi\sum_{\eta\eta'\lambda}\sum_{{\bf kk'}}n_i|V_{\bf kk'}|^2\eta'\eta\lambda\frac{(4\Delta_0^2-\Omega^2)\xi_{\bf k}^2}{4E_{\bf k}^3\Lambda_{\bf k'}(E_{\bf k'}^{\lambda}\!-\!\Omega)}\frac{f(E_{\bf k}^{\eta}\!-\!\Omega)\delta(E_{\bf k}^{\eta}\!-\!E_{\bf k'}^{\lambda})-f(E_{\bf k'}^{\lambda}\!-\!\Omega)\delta(E_{\bf k}^{\eta'}\!+\!\Omega\!-\!E_{\bf k'}^{\lambda})}{E_{\bf k}^{\eta}\!-\!\Omega\!-\!E_{\bf k}^{\eta'}}\nonumber\\
&\approx&-{\delta|\Delta|^2}i\pi\sum_{\eta\eta'\lambda}\sum_{{\bf kk'}}n_i|V_{\bf kk'}|^2\eta'\eta\lambda\frac{(4\Delta_0^2-\Omega^2)\xi_{\bf k}^2}{4E_{\bf k}^3\Lambda_{\bf k'}(E_{\bf k'}^{\lambda}\!-\!\Omega)}\frac{f(E_{\bf k}^{\eta}\!-\!\Omega)\delta(E_{\bf k}^{\eta}\!-\!E_{\bf k'}^{\lambda})-f(E_{\bf k'}^{\lambda}\!-\!\Omega)\delta(E_{\bf k}^{\eta'}-\!E_{\bf k'}^{\lambda})}{E_{\bf k}^{\eta}\!-\!\Omega\!-\!E_{\bf k}^{\eta'}}.
\end{eqnarray}
The $\eta'=\eta$ part vanishes in above equation. Then, with $\eta'=-\eta$, one has 
\begin{eqnarray}
  H_{\rm vc}&=&-{\delta|\Delta|^2}i\pi\sum_{\eta\lambda}\sum_{{\bf kk'}}n_i|V_{\bf kk'}|^2\lambda\frac{(4\Delta_0^2-\Omega^2)\xi_{\bf k}^2}{4E_{\bf k}^3\Omega(\Omega-2\lambda{E_{\bf k'}})}\Big[\frac{f(E_{\bf k}^{\eta}\!-\!\Omega)}{\Omega\!-\!2{\eta}E_{\bf k}^{\eta}}\!+\!\frac{f(E_{\bf k'}^{\lambda}\!-\!\Omega)}{\Omega\!+\!2{\eta}E_{\bf k}^{\eta}}\Big]\delta(E_{\bf k}^{\eta}\!-\!E_{\bf k'}^{\lambda})\nonumber\\
  &=&-{\delta|\Delta|^2}i\pi\sum_{\eta}\sum_{{\bf kk'}}n_i|V_{\bf kk'}|^2\frac{(4\Delta_0^2-\Omega^2)\xi_{\bf k}^2}{2E_{\bf k}^3(\Omega-2\eta{E_{\bf k}})}\frac{{\eta}f(E_{\bf k}^{\eta})}{\Omega^2\!-\!4E_{\bf k}^2}\delta(E_{\bf k}\!-\!E_{\bf k'}) \nonumber\\
  &\approx&-{\delta|\Delta|^2}i\Omega\pi\sum_{{\bf kk'}}n_i|V_{\bf kk'}|^2\frac{\Delta_0^2\xi_{\bf k}^2}{4E_{\bf k}^7}\frac{2f(E_{\bf k})\!-\!1}{2}\delta(E_{\bf k}\!-\!E_{\bf k'})-{\delta|\Delta|^2}i\pi\sum_{{\bf kk'}}n_i|V_{\bf kk'}|^2\frac{\Delta_0^2\xi_{\bf k}^2}{4E_{\bf k}^6}\delta(E_{\bf k}\!-\!E_{\bf k'}). \label{Hvc}
\end{eqnarray}
It is noted that the last term on the right-hand side of above equation is irrelevant to the Higgs-mode damping and can be neglected.  Consequently, with Eqs.~(\ref{Hba}) and (\ref{Hbc}) as well as (\ref{Hvc}), from the nonequilibrium action  $\delta{S^{(2)}}=-H_{\rm ba}-H_{\rm bc}-H_{\rm vc}-\delta|\Delta|^2/g$, Eq.~(\ref{HNS}) is derived.

\end{appendix}

\end{widetext}


\begin{thebibliography}{0}
\bibitem{Silaev1}  It is noted that within the Eilenberger equation, Ref.~\onlinecite{Silaev} derives a vanishing Higgs-mode generation in the second-order optical response at clean limit. Although the derivation in Refs.~\onlinecite{Cea1,Cea2,Cea3} through path-integral approach also find the same result, this result holds strongly  against the Ginzburg-Landau theory, which reveals an opposite conclusion (i.e., a finite Higgs-mode generation in the second-order optical response at clean limit)\cite{EPM}.  It is further analytically proved\cite{EPM} that both derivations in Ref.~\onlinecite{Silaev} and Refs.~\onlinecite{Cea1,Cea2,Cea3} contain the mathematical flaws. After fixing these flaw, a finite Higgs-mode generation through the drive effect of vector potential is derived at clean limit\cite{EPM}, exactly recovering the previous result from GIKE\cite{GIKE2} as well as Ginzburg-Landau theory.
\bibitem{Silaev0} It is noted that Ref.~\onlinecite{Silaev} also used the diagrammatic formalism in order to justify/recover the result of the scattering from the Eilenberger equation. Nevertheless, a special procedure, by first taking the commutation between the self-energy and $\tau_3$-Green function and applying the quasiclassical approximation afterwards, was applied to handle the calculation within the diagrammatic formalism (refer to Sec.~IV~A in Ref.~\onlinecite{Silaev}). This procedure is not the standard diagram formulation\cite{G1}, but follows the exact step to derive the Eilenberger equation\cite{Eilen1}. Consequently, the applied diagrammatic formalism approach with this procedure is exactly same as the Eilenberger equation rather than a separate nontrivial approach.
\end{thebibliography}
\end{document}